\documentclass[sts,preprint]{imsart}

\usepackage[english]{babel} 
\usepackage{amssymb,amsmath,amsthm}
\usepackage{txfonts}
\usepackage{mathdots}
\usepackage[classicReIm]{kpfonts}
\usepackage[dvips]{graphicx} 
\usepackage{longtable} 
\usepackage{tabu} 
\usepackage[table]{xcolor}
\usepackage{adjustbox} 
\usepackage{enumitem} 

\usepackage{tikz} 
\usetikzlibrary{shapes,arrows}
\tikzstyle{mynode} = [rectangle, rounded corners, minimum width=3cm, minimum height=1cm,text centered, draw=black]
\tikzstyle{arrow} = [thick,->,>=stealth]

\usepackage{natbib}

\RequirePackage[colorlinks,citecolor=blue,urlcolor=blue]{hyperref} 

\startlocaldefs
\definecolor{tableHeader}{RGB}{211, 47, 47}
\definecolor{tableLineOne}{RGB}{245, 245, 245}
\definecolor{tableLineTwo}{RGB}{224, 224, 224}

\endlocaldefs

\begin{document}
\begin{frontmatter}
	
	\title{A Review of Generalizability and Transportability}
	\runtitle{Generalizability and Transportability}
	
	
	\author{\fnms{Irina} \snm{Degtiar}\corref{}\ead[label=e1]{idegtiar@g.harvard.edu}}
	\and
	\author{\fnms{Sherri} \snm{Rose}\ead[label=e2]{sherrirose@stanford.edu}}
	\affiliation{Harvard T.H. Chan School of Public Health and Stanford University}
	
	\address{Irina Degtiar is a PhD candidate at the Department of Biostatistics, Harvard T.H. Chan School of Public Health, 655 Huntington Ave, Boston, MA 02115, USA (email: \href{mailto:idegtiar@g.harvard.edu}{idegtiar@g.harvard.edu}).}
	
	\address{Sherri Rose is an Associate Professor at the Center for Health Policy and Center for Primary Care and Outcomes Research, Stanford University, 615 Crothers Way, CA 94305, USA (email: \href{mailto:sherrirose@stanford.edu}{sherrirose@stanford.edu}).}
	
	\runauthor{I. Degtiar et al.}
	
	\begin{abstract}
	When assessing causal effects, determining the target population to which the results are intended to generalize is a critical decision. Randomized and observational studies each have strengths and limitations for estimating causal effects in a target population. Estimates from randomized data may have internal validity but are often not representative of the target population. Observational data may better reflect the target population, and hence be more likely to have external validity, but are subject to potential bias due to unmeasured confounding. While much of the causal inference literature has focused on addressing internal validity bias, both internal and external validity are necessary for unbiased estimates in a target population. This paper presents a framework for addressing external validity bias, including a synthesis of approaches for generalizability and transportability, the assumptions they require, as well as tests for the heterogeneity of treatment effects and differences between study and target populations. 
	\end{abstract}
	
	
	
\end{frontmatter}

\section{Background}

The goal of causal inference is often to gain understanding of a particular target population based on study findings. The true underlying causal effect will typically vary with the definition of the chosen target population. However,  samples unrepresentative of the target population arise frequently in studies ranging from randomized controlled trials (RCTs) in clinical medicine to policy research \citep{bell2016,kennedy-martin2015,allcott2015}.  
In a clinical trial setting, physicians may be left interpreting evidence from RCTs with patients who have demographics and comorbidities that are quite different from those of their patients.
As an example, within cancer RCTs, African Americans are widely underrepresented despite being at an increased risk for many cancers \citep{chen2018}. Failing to address this lack of representation can lead to inappropriate conclusions and harm \citep{chen2020}. In a policy setting, it is important to consider the effects that can be expected in the eventual target population in order to set expectations for anticipated results and determine groups that should be targeted for an intervention.
 
  \begin{figure} 
	\caption[]{Internal vs. external validity biases as they relate to target, study, and analysis populations.} 
	\label{Figure_internal_external_validity}
	\includegraphics*[width=1\textwidth, keepaspectratio=true]{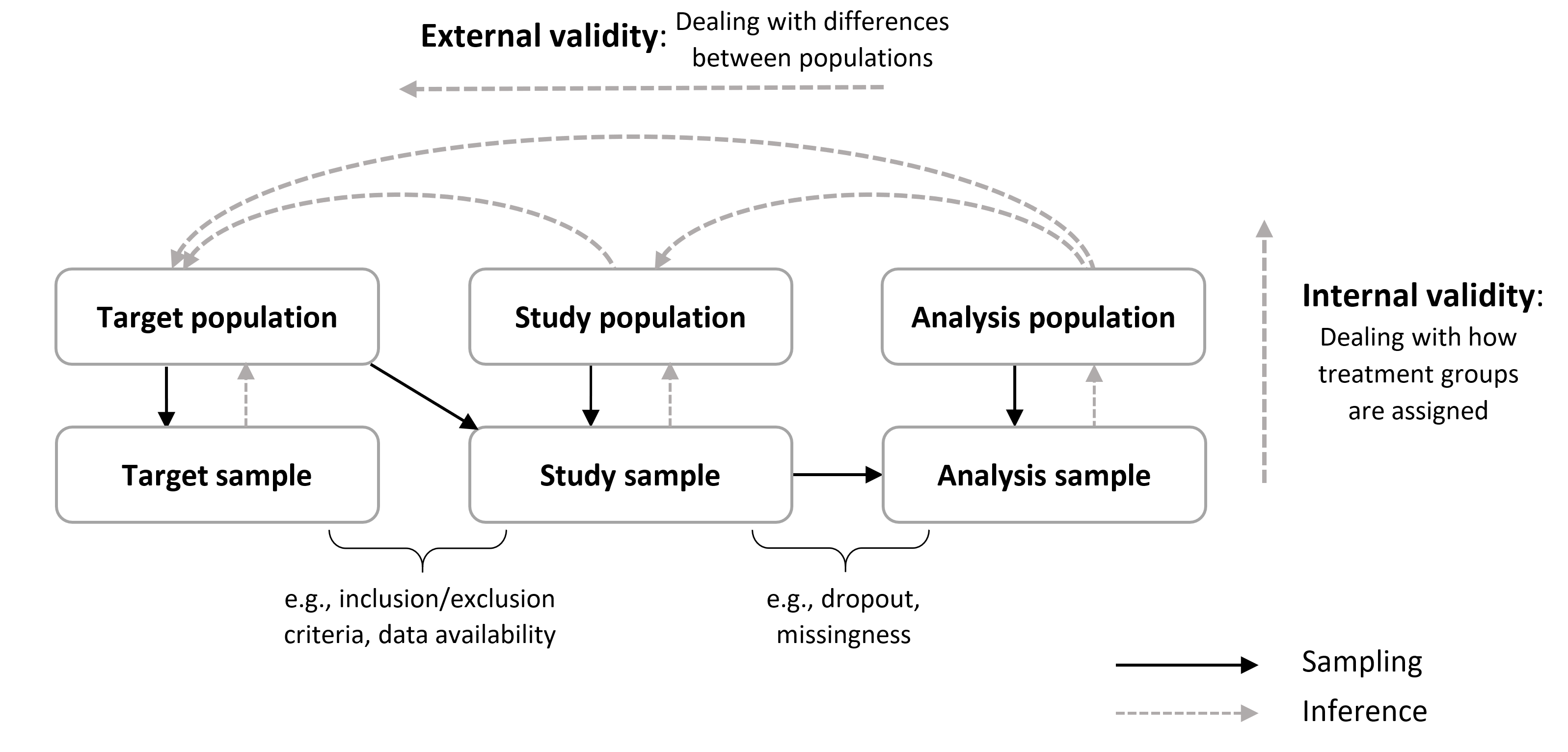}
\end{figure}

The relationships between target, study, and analysis populations are visualized in Figure~\ref{Figure_internal_external_validity}. The target sample is a representative sample of the target population, whereas the study population is defined by enrollment processes and inclusion or exclusion criteria. Due to these practical and scientific considerations, the study population may differ from the target population. Correspondingly, the enrolled participants who form the study sample may have different characteristics from those of the target sample. In the cancer RCT example, while a physician might care about the target population of patients that may come in to be treated by their clinic (of which the clinic's current patients are a target sample), the study sample on which they're basing their treatment recommendations may not include any African Americans. The study population is the hypothetical population that the study sample represents, which likewise includes no African Americans. Post-enrollment, further dropout and missingness may occur that create the observed analysis sample. In this case, dropout may have occurred for patients who experienced severe adverse events such that the analysis sample consists of patients who did not experience severe side effects. There then exists a hypothetical analysis population from which the  analysis sample data is a simple random sample. Hereafter, for simplicity and consistency with the literature, we will use the terms study sample and study population to be inclusive of the analysis sample and analysis populations, respectively.

Several key concepts are crucial to understand when considering extending causal inferences beyond a study sample. \textit{Generalizability} focuses on  the setting where the study population is a subset of the target population of interest, while \textit{transportability}  addresses the setting where the study population is (at least partly) external to the target population.  \textit{Internal validity} is defined as an effect estimate being unbiased for the causal treatment effect in the population from which the sample is a simple random sample (i.e., moving vertically from a sample to its corresponding population in Figure~\ref{Figure_internal_external_validity}).  \textit{External validity} is concerned with how well results generalize to other contexts. Specifically, that the (internally valid) effect estimate is unbiased for the causal treatment effect in a different setting, such as a target population of interest (moving laterally between populations in Figure~\ref{Figure_internal_external_validity}). External validity bias has also been referred to as sample selection bias \citep{heckman1979,imai2008,moreno-torres2012,bareinboim2014,haneuse2016}.

External validity bias arises from differences between the study and target populations in (1) subject characteristics; (2) setting, such as geography or type of health center; (3) treatment, such as timing, dosage, or staff training; and (4) outcomes, such as length of follow-up or timing of measurements \citep{cronbach1982,rothwell2005,dekkers2010,green2006,burchett2011,attanasio2003}. The focus
of most generalizability and transportability methods is on addressing differences
in subject characteristics. Hence, these methods assume the remaining
threats to external validity are not present in the data sources they are looking to
generalize across. Namely, external validity bias then arises solely from: (1) variation in the probability of enrollment in the study, (2) heterogeneity in treatment effects, and (3) the correlation between (1) and (2)  \citep{olsen2013}.  We  therefore distinguish between factors differentiating the target population from the study population (external validity bias) and those that create differences between treatment groups (internal validity bias), e.g., confounding.  RCTs are frequently performed in a nonrepresentative subset of the target population and may have imperfect follow-up (challenging their external validity) and may have baseline imbalances (leading to internal validity bias). Observational studies may be susceptible to unmeasured confounding (threatening their internal validity), but may be more representative of the target population (hence having better external validity).  Lack of representation in an RCT can lead to external validity bias that is  larger than the internal validity bias of an observational study \citep{bell2016}.

  The optimal solution to external validity bias centers on study design, which we review briefly here, but do not cover extensively. One type of ideal study would  randomly sample subjects from the target population and then randomly assign treatment to the selected individuals. However, this  is usually infeasible. Alternative study designs for improving study generalizability and transportability include purposive sampling, where investigators deliberately select individuals such as for representation or heterogeneity \citep{shadish2001,allcott2012}; pragmatic or practical clinical trials, which aim to be representative of clinical practice \citep{schwartz1967,ford2016}; stratified selection based on effect modifiers or propensity scores for selection \citep{tipton2014a,tipton2013a,allcott2012}; and balanced sampling designs for site selection that select representative sites through stratified ranked sampling \citep{tipton2017}.   In lieu or in addition to study designs that address external validity bias, generalizability and transportability methods can improve the external validity of effect estimates after data collection.
 
This manuscript  provides a review of generalizability and transportability research, synthesizing across the statistics, epidemiology, computer science, and economics literature in a more complete manner than has been done to date. 
Existing review literature has examined  narrower subsets of the topic: generalizing or transporting  to a target population from only RCT data \citep{stuart2015,stuart2018,kern2016,tipton2018,ackerman2019}, identifiability rather than estimation \citep{bareinboim2016}, or meta-analysis approaches for combining summary-level information \citep{verde2015,kaizar2015}.  A recent related review on combining randomized and observational data featured a simulation, real data analysis, and software guide \citep{colnet2020}.
However, these previous reviews have not summarized the full range of generalizability and transportability methods that incorporate  data from randomized, observational, or a combination of randomized and observational studies, nor techniques for evaluating generalizability, as we do here.   Additionally,  although the importance of describing generalizability and transportability is recognized by different trial reporting guidelines (e.g., CONSORT, RECORD, STROBE), they provide no clear guidance on tests or estimation procedures \citep{schulz2010,benchimol2015,vonelm2008}. We also contribute recommendations for methodologists and applied researchers.

\begin{figure}
	\caption[]{Overview framework for assessing and addressing external validity bias after data collection.}
	\label{Figure_framework}
	\begin{tikzpicture}[node distance=1.75cm]
		\node (n1) [mynode, text width=12cm] {\textit{Estimand}: consider study and target populations, and with them, the estimand of interest};
		\node (n2) [mynode, below of=n1, text width=12cm] {\textit{Assumptions}: assess validity of assumptions necessary for generalizability or transportability approaches};
		\node (n3) [mynode, below of=n2, text width=12cm] {\textit{Evaluating Generalizability}: examine whether treatment effect modification exists and whether effect modifiers differ in distribution between study and target populations};
		\node (n4) [mynode, below of=n3, text width=12cm] {\textit{Generalizability and Transportability Methods}: \\apply  methods for addressing external validity bias};
		
		\draw [arrow] (n1) -- (n2);
		\draw [arrow] (n2) -- (n3);
		\draw [arrow] (n3) -- (n4);
	\end{tikzpicture}
\end{figure}
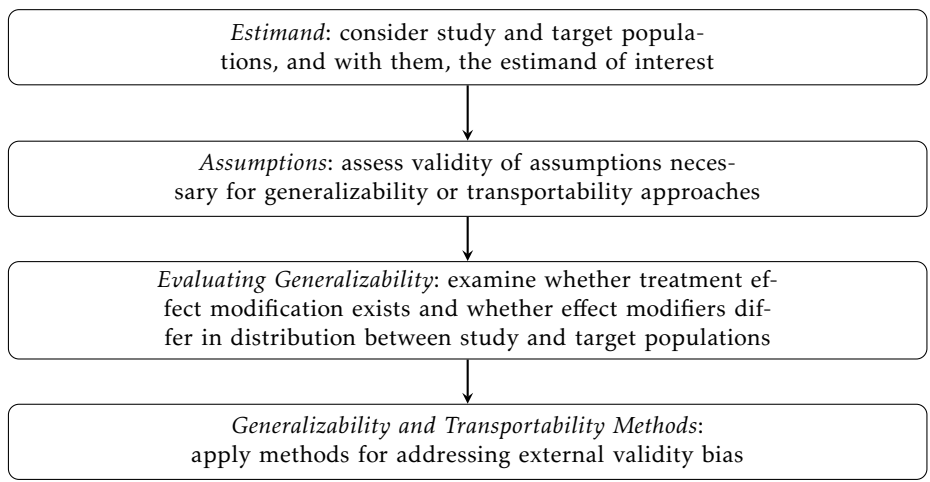
 
 The remainder of the article synthesizes considerations for assessing and addressing external validity bias after data collection (presented as a framework in Figure 2) and is organized as follows.  
 Section 2 defines the estimand of interest, the average treatment effect in a target population, as well as alternatives. Section 3 presents key assumptions underlying many of the  methods. Section 4 reviews methods for assessing treatment effect heterogeneity, thus further motivating the need for methods that enable generalizing or transporting study results to a target population. Section 5 then summarizes  the analytic methods available for external validity bias correction that generate treatment effect estimates for a target population of interest. These techniques include weighting and matching, outcome regressions, and doubly robust approaches. Section 6 then concludes with guidance for both applied and methods researchers.


\section{Estimand}

Assume, for one or more studies, the existence of outcome $Y$, treatment $A \in \{0,1\}$, and baseline covariates ${X} \in \mathbb{R}^d$. For simplicity of notation, we define \textit{X} to represent all treatment effect confounders and effect modifiers (subgroups whose effects are expected to differ) that differ between study and target populations; each variable in \textit{X} is both a confounder and an effect modifier. Without loss of generality, we focus on the single study setting, with $S=1$ indicating selection into it.
The observational unit for the study sample is $O_{\text{study}} = \{X,A,Y,S=1\}$. $O_{\text{study}}$ has probability distribution $P_{\text{study}}\in \cal{M}_{\text{study}}$, where $\cal{M}_{\text{study}}$ is our collection of possible probability distributions (i.e., statistical model). We observe $n_s$ realizations of $O_{\text{study}}$, indexed by $j$.
The observational unit for a representative sample from the target population is given by $O= \{X,A,Y,S\}\sim P \in \cal{M}$. We observe $n$ realizations of $O$, indexed by $i$. Target sample subjects who do not appear in the study sample will have $S=0$. We use the terminology ``selected'' or ``sampled'' throughout the paper for simplicity although for transportability, subjects are not directly sampled into the study from the target population. 
For generalizability, $O_{\text{study}} \in O$, while for transportability, the two are disjoint sets, $O_{\text{study}} \notin O$.
 
Biases are defined with respect to an estimand. We will focus on the average treatment effect in a well-defined target population of interest: the  population average treatment effect (PATE). Namely, we are interested in the average outcome had everyone in the target population been assigned to treatment \textit{A=1} compared to the outcome had everyone been assigned to treatment \textit{A=0}. We write this as $\tau=E_X(E(Y|S=1,A=1,X)-E(Y|S=1,A=0,X))=E(Y^1-Y^0)$, where $Y^1$ and $Y^0$ are the potential outcomes under treatment and no treatment, respectively, and required identifiability assumptions are delineated in the next section. The corresponding estimator is given by $\hat{\tau}=1/{n}\sum^{n}_{i=1}{(\hat{Y}^1_i-\hat{Y}^0_i)}$. We also write $Y^a$ to represent the potential outcome under $a$ with lowercase $a$ a specific value for random variable $A$.  Potential outcomes are either explicitly assumed in the potential outcomes framework or a consequence of the structural causal model \citep{rubin1974,pearl2000}. Different target populations correspond to alternative PATEs because the expectation is taken with respect to alternative distributions of covariates $X$. However, necessarily, we only observe outcomes in the study sample. A study therefore directly estimates the  sample average treatment effect (SATE): $\tau_s=E(Y^1-Y^0|S=1)$ with estimator $\hat{\tau}_s=1/n_s\sum_{j:S_j=1}{(\hat{Y}^1_j-\hat{Y}^0_j)}$.

\begin{figure} 
	\caption[]{Illustrative example of the difference between target population and sample average treatment effects (PATE and SATE). Biases may differ in magnitude and may make the SATE either larger or smaller than the PATE.}
	\label{Figure_SATE_PATE}
	\includegraphics*[width=0.7\textwidth, keepaspectratio=true]{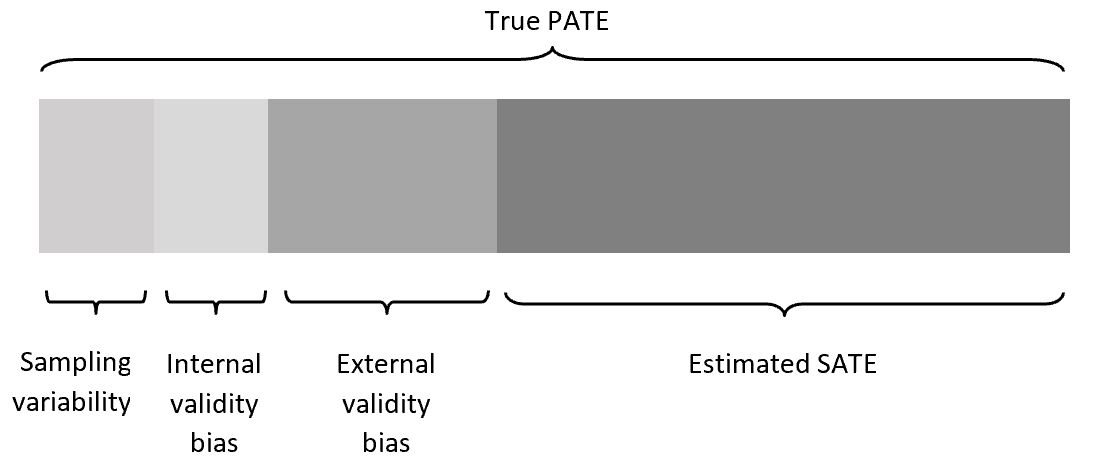}
\end{figure}

When the distributions of treatment effect modifiers differ between study and target populations, the true study average effect will not equal the true target population average effect (SATE $\neq $ PATE) due to external validity bias. Sampling variability as well as internal validity biases can also drive estimates of SATE further from the truth (Figure~\ref{Figure_SATE_PATE}). Biases may differ in magnitude and may make the SATE either larger or smaller than the PATE.

We may also be interested in estimating other target parameters.  For example, the population conditional average treatment effects (PCATE): $\tau_x=E(Y^1-Y^0|X)$ is examined in some of the estimation methods we explore later. Another parameter of interest is the population average treatment effects among the treated: $\tau_1=E(Y^1-Y^0|A=1)$. Similar generalizability and transportability considerations presented in the following sections will apply for these and other causal estimands.

\section{Assumptions}

Under the potential outcomes framework, the assumptions below are sufficient to identify the PATE using the observed study data. A corresponding set of assumptions under the structural equation model (SEM) framework has also been derived \citep{pearl2014,pearl2015,pearl2011,bareinboim2014a,bareinboim2015,bareinboim2016,correa2018}.  Additional  assumptions include those of no missing data or measurement error in outcome, treatment, or covariate measurements. Other target parameters of interest necessitate a similar set of assumptions.

\subsection{Internal validity}

\noindent  Sufficient assumptions for identifying the PATE with respect to internal validity:

 \textbf{Conditional treatment exchangeability}: $Y^a\bot A\ |X,S=1$ for all $a\in \mathcal{A}$, the set of all possible treatments. This condition requires no unmeasured confounding of the treatment-outcome relationship in the study. It is satisfied by perfectly randomized trials (e.g., no loss to follow-up, other informative missingness or censoring, etc.) and by observational studies that have all confounders measured. 
 While this condition is sufficient, it is not always necessary. When estimating the PATE, it can be replaced by the weaker condition of mean conditional exchangeability of the treatment effect, $E(Y^1-Y^{0}|X,A,S=1)=E(Y^1-Y^{0}|X,S=1)$ \citep{kern2016,dahabreh2019a}.

 \textbf{Positivity of treatment assignment}: $P(X=x|S=1)>0\ \Rightarrow P(A=a|X=x,$ $S=1)>0$, with probability 1 for all $a\in \mathcal{A}$. This condition entails that each subject in the study has a positive probability of receiving each version of the treatment. In combination with the conditional treatment exchangeability assumption above, this assumption is also known as strongly ignorable treatment assignment \citep{varadhan2016}. 

 \textbf{Stable unit treatment value assumption (SUTVA)}: if $A=a$ then $Y=Y^a$. This assumption requires no interference between subjects and treatment version irrelevance (i.e., consistency/well-defined interventions) in the study and target populations \citep{dahabreh2017,kallus2018}.

\subsection{External validity}

\noindent Following the assumptions above, identifying the PATE involves a parallel set of assumptions for external validity:

 \textbf{Conditional exchangeability for study selection}: $Y^a\bot S\ |X$ for all $a\in $ $\mathcal{A}$. This assumption is also known as exchangeability over selection and the generalizability assumption. It requires that the outcomes among individuals with the same treatment and covariate values in the study and target populations are the same \citep{stuart2011}. All effect modifiers that differ between study and target populations must therefore be measured. This assumption would be satisfied by a study sample that is a random sample from the target population or a nonprobability study sample in which all effect modifiers are measured. A weaker condition, mean conditional exchangeability of selection, $E(Y^1-Y^{0}|X,S=1)=E(Y^1-Y^{0}|X)$ can replace conditional exchangeability for study selection when focusing on the PATE \citep{kern2016,dahabreh2019a}.

 \textbf{Positivity of selection}: $P\left(X=x\right)>0\ \Rightarrow P\left(S=1|X=x\right)>0$ with probability 1 for all $a\in \mathcal{A}$. This assumption requires common support with respect to study selection; in every stratum of effect modifiers, there is a positive probability of being in the study sample \citep{dahabreh2017}. This can be replaced by smoothing assumptions under a parametric model, for example, that the propensity score distribution has sufficient overlap or common support between the study sample and target population \citep{westreich2017,tipton2017a}. Thus, with conditional positivity of selection we assume that all members of the target population are represented by individuals in the study. The positivity assumption in combination with the no unmeasured effect modification assumption above is also known as strongly ignorable sample selection given the observed covariates \citep{chan2017}.

 \textbf{SUTVA for study selection}: if $S=s$ (and $A=a$) then $Y=Y^a$. This assumption states that there is no interference between subjects selected into the study versus those not selected and that there is treatment version irrelevance between study and target samples (the same treatment is given to both) \citep{tipton2013,tipton2017a}. It necessitates no difference across study and target samples in how outcomes are measured or in how the intervention is applied, that there is a common data-generating function for the outcome across individuals in the study and target populations (i.e., that being in the study does not change treatment effects), and that the potential outcomes are not a function of the proportion of individuals selected for the study. Treatment version irrelevance in SUTVA can be replaced by the  condition of having the same distribution of treatment versions between study and target populations when estimating the PATE \citep{lesko2017}.

\subsection{Transportability}

 Similar internal and external validity assumptions are needed for transportability, with the following modifications. When the study sample is a subset of the target population (generalizability), the positivity assumption for selection will need the propensity for selection to be bounded away from 0, whereas when the sample is not a subset of the target population (transportability), the propensity to be in the target population will need to be bounded away from 0 and 1 \citep{tipton2013}. Furthermore, for transportability, the set of covariates, \textit{X}, required for conditional exchangeability for study selection cannot include those that separate the study sample from the target population (e.g., hospital type if transporting results from teaching hospitals to community clinics, or geographic location if transporting between states) \citep{tipton2013}. Further distinctions are discussed by \cite{pearl2015} using the SEM framework. Under this framework, Pearl and Bareinboim formalize the assumptions necessary for using different transport formulas to reweight randomized data, providing graphical conditions for identifiability as well as transport formulas for randomized studies \citep{pearl2014,pearl2015}, observational studies \citep{pearl2011,pearl2015,bareinboim2015,bareinboim2016,correa2017,correa2018}, and a combination of heterogeneous studies \citep{bareinboim2014a,bareinboim2016}.

\section{Assessing dissimilarity between target and study populations and testing for treatment effect heterogeneity}

Numerous quantitative approaches can help evaluate the extent to which study results may be expected to generalize to the target population. These assessments examine population differences and whether treatment effect heterogeneity exists.
 Methods for assessing the similarity of study and target populations can broadly be categorized into those that compare baseline patient characteristics and those that compare outcomes for groups on the same treatment. For the former, many make use of the propensity score for selection, which also serves the purpose of assessing the extent to which propensity score adjustment using measured covariates can sufficiently remove baseline differences between study and target samples. However, most of these methods do not emphasize effect modifiers, hence should be combined with an assessment of whether the noted population differences correspond to heterogeneity of treatment effects.  To test for heterogeneity of effects, one must first identify effect modifiers. Effect modifiers are often pre-specified by the investigator, but data-driven approaches exist as well, and will be discussed in this section.

\subsection{Assessing dissimilarity between populations using baseline  characteristics}

When summary-level study data are available,  assessments that examine differences in univariate covariate metrics between study and target samples can be deployed. \cite{cahan2017} propose a generalization score for evaluating clinical trials that incorporates baseline patient characteristics, the trial setting, protocol, and patient selection: it takes ratios of the mean or median values of these characteristics in the study and target samples, then averages across categories for an overall score. However, this approach does not account for any measures of dispersion, which may reflect exclusion of more heterogeneous individuals from the study. When only baseline patient characteristics are responsible for relevant study vs. target population differences, one can perform multiplicity-adjusted univariate tests for differences in effect modifiers between study and target samples \citep{greenhouse2008}. Alternatively, one could examine absolute standardized mean differences (SMD) for each covariate, $(\bar{X}_{\text{study}}-\bar{X})/\sigma_{\bar{X}}$, where $\bar{X}_{\text{study}}$ and $\bar{X}$ are the means of baseline covariates in the study and target samples, respectively, and $\sigma_{\bar{X}}$ is the standard deviation of $\bar{X}$ \citep{tipton2017a}.
High values indicate heavy extrapolation and reliance on correct model specification; in  smaller samples, imbalances will often occur by chance \citep{tipton2017a}. With one or more RCTs, generalizability across categorical eligibility criteria can be assessed by the percent of the target sample that would have been eligible for the study or set of studies \citep{weng2014,he2016,sen2016}.

 Joint distributions of patient characteristics can likewise be compared, such as by examining the SMD in propensity scores for selection 
 \citep{stuart2011}. When the propensity score is not symmetrically distributed, summarizing mean differences is insufficient. \cite{tipton2014} developed a generalizability index that bins propensity scores and  is bounded between 0 and 1: $\sum_{j=1}^k \sqrt{w_{p_j} w_{s_j}}$ with $j=1,...,k$ bins, each with target sample proportions $w_{p_j}$ and study sample proportions $w_{s_j}$. It is based on the distributions of propensity scores rather than only the averages. However, this approach requires patient-level study and target sample data.  
 A generalizability index score of $\mathrm{<}$0.5 indicates a study being very challenging to generalize from and a score of $\mathrm{>}$0.9 indicates high generalizability \citep{tipton2014}. 
 Other propensity score distance measures can be used, such as Q-Q plots, Kolmogorov-Smirnov distance, Levy distance, the overlapping coefficient, and C statistic; these largely focus on comparing cumulative densities \citep{tipton2014,ding2016}. To assess the degree of extrapolation with respect to effect modifiers, one can examine overlap in the propensity of selection distributions, such as the proportion of target sample individuals with propensity scores outside the 5${}^{th}$ and 95${}^{th}$ percentiles of the sample propensity scores \citep{tipton2017a}.
 
 One can also adopt a machine learning approach for detecting covariate shift--a change in the distribution of covariates between training and test data (here, the study and target data) \citep{glauner2017}. After creating a joint dataset with target and study sample data, a classification algorithm predicts whether the data came from the study. A dissimilarity metric surpassing a threshold of acceptability then indicates sizable dissimilarity between datasets. 
 However, an inability to accurately predict study vs. target data origin does not rule out differences in effect modifiers. A low score might furthermore indicate an incorrect model specification or insufficient model tuning.

  The tests discussed in this subsection assess differences between populations; however, they require investigator knowledge of which characteristics moderate the treatment effect (or are correlated with unmeasured effect modifiers) and what level of differences are clinically relevant. Many covariates are often tested 
 or included in a propensity score regression for study selection. This approach prioritizes predictors that are strongly associated with study selection rather than those that exhibit strong effect modification. Investigators should therefore aim to identify relevant effect modifiers for testing or inclusion in the propensity score regression and test this subset.

\subsection{Assessing dissimilarity between populations using outcomes}
\label{subsection:test_dissimilarity}

When individual-level outcome data or joint distributions of group-level outcome data are available in both the study and target samples for at least one of the treatment groups, the following methods can assess the extent to which measured effect modifiers account for population differences.  One can compare the observed outcomes in the target sample to predicted outcomes using study controls \citep{stuart2011}, or more generally, study individuals who received the same treatment \citep{hotz2005}:  $1/n_a\sum_{i=1}^N 1(A_i=a) Y_i$ vs. $1/n_{s,a}\sum_{i:S_i=1}1(A_i=a)w_i Y_i$ with weights $w_i$ defined by weighting and matching methods discussed in Section \ref{weighting_matching}. \cite{hartman2015} formalize this comparison with equivalence tests.
Alternatively,  conditional outcomes for study and non-study target sample individuals receiving the same treatment, conditioning on measured effect modifiers, can be compared to detect unmeasured effect modification, although other identifiability assumption violations might also be at fault: $E(Y|X,A=a,S=1)$ vs. $E(Y|X,A=a,S=0)$. Possible tests include analysis of covariance, Mantel-Haenszel, U-statistic based tests, stratified log-rank, or stratified rank sum, depending on the outcome \citep{marcus1997,hotz2005,luedtke2019}. For example, study controls could be compared to subgroups of the target population that were known to be excluded from the study (e.g., patients who declined participation in a RCT, as done by \cite{davis1988}). Relatedly, unmeasured effect modification can be imperfectly tested for by disaggregating a characteristic that differentiates the study from the target sample \citep{allcott2012}. These outcome differences should not exceed those observed between study treatment groups \citep{begg1992}. 

In addition to testing for outcome differences, one can test for differences between study and target regression coefficients or between baseline hazards in a Cox regression \citep{pan2009}. Any identified differences in outcomes or effects will reflect sample differences unaccounted for by the outcome or weighting method, indicating unmeasured effect modification or an ineffective modeling approach. To have this comparison reflect relevant differences, study controls must be representative of the target population after weighting or regression adjustment. 
\cite{hartman2015} provides a more formal set of identifiability assumptions that may be violated when each equivalence test is rejected. If unmeasured effect modification is suspected, one can perform sensitivity analysis to assess the extent to which it can impact results \citep{marcus1997,nguyen2017,nguyen2018,dahabreh2019c,andrews2017} or to generate bounds on the treatment effect when only partial identification is possible \citep{chan2017}.

\subsection{Testing for treatment effect heterogeneity }\label{TEH}

 Identified population differences are relevant insofar as they correspond to differences in treatment effect modifiers. The following tests enable an investigator to assess whether treatment effects vary substantially across measured covariates. Many are suitable for use in observational or RCT data, although have largely been demonstrated in RCT data to date. While some tests require a priori specification of subgroups, others can discover them in data-driven ways and most require individual-level data. A straightforward, but often overlooked issue is that studies with enrolled patients that are homogeneous with respect to effect modifiers will have difficulty identifying heterogeneity of effects. These approaches are therefore best applied to data representative of the target populations \citep{gunter2011}.

 Tests of prespecified subgroups should focus on target population subgroups under- or over-represented in the study, or any other clinically relevant subgroup expected to exhibit effect heterogeneity. Largely,  methods for testing treatment effect heterogeneity of a priori specified subgroups exhibit limited power. Those testing several effect modifiers individually are particularly underpowered to detect significant effects once  multiple testing adjustments are incorporated. One approach tests the interaction term of treatment assignment with an effect modifier in a linear model, which also requires modeling assumptions as to the linearity and additivity of effects \citep{fang2017,gabler2009}. To address this lack of power, sequential tests for identifying treatment-covariate interactions can be used with either randomized or observational data \citep{qian2019}. Alternative approaches, each addressing slightly different goals, include testing whether the conditional average treatment effect is identical across predefined subgroups \citep{crump2008,green2012}, comparing subgroup effects to average effects \citep{simon1982}, and identifying qualitative interactions or treatment differences exceeding a prespecified clinically significant threshold \citep{gail1985}. 

When effect modifiers are not known a priori, a variety of techniques can be applied for identifying subgroups with heterogeneous effects. These include those that identify variables that qualitatively interact with treatment (i.e., for which the optimal treatment differs by subgroup) \citep{gunter2011} as well as determine the magnitude of interaction \citep{chen2017,tian2014}. Various machine learning approaches can also be used to identify subgroups with heterogeneous treatment effects while minimizing modeling assumptions. 
Approaches that also present tests for treatment effect differences between subgroups include Bayesian additive regression trees (BART) and other classification and regression tree (CART) variants \citep{su2008,su2009,lipkovich2011,green2012,athey2016}. Tree-based methods  develop partitions in the covariate space recursively to grow toward terminal nodes with homogeneity for the outcome.    
These approaches may be particularly useful when heterogeneity may be a function of a more complex combination of factors.

 With many effect modifiers or when effect modifiers are unknown, global tests for heterogeneity can also be used. \cite{pearl2015} provides conditions for identifying treatment effect heterogeneity (including heterogeneity due to unmeasured effect modifiers) for randomized trials with binary treatments, situations with no unobserved confounders, and with mediating instruments. 
 Effect heterogeneity can  be tested for using the baseline risk of the outcome as an effect modifier; interaction-based tests assess for differences in baseline risk between study and target population control groups \citep{varadhan2016,weiss2012}. These tests avoid the need for multiple testing but require outcome data in the target sample and modeling assumptions. A consistent nonparametric test also exists that assesses for constant conditional average treatment effects, $\tau_x = \tau\ \forall x \in \mathcal{X}$ \citep{crump2008}. Additional methods, which suffer from limited power and rely on estimates of SATE, include testing whether potential outcomes across treatment groups have equal variances and whether cumulative distribution functions of treatment and control outcomes differ by a constant shift \citep{fang2017}. Global tests do not identify subgroups responsible for effect heterogeneity, although if a global test is rejected, one can then compare individual subgroups to determine which demonstrate effect heterogeneity.

 If these assessments of generalizability fail and the target population is not well-represented by the study population (specifically, when strong ignorability fails), \cite{tipton2013} provides several recommended paths forward. Investigators can change the target population to one represented by the study. That is, change the estimand of interest by aligning inclusion and exclusion criteria, outcome timepoints, or treatment doses \citep{hernan2008,weisberg2009}. A population coverage percentage can then summarize the percent overlap between the new and original target sample propensity scores, and describe relevant differences from the original target population. Investigators can alternatively retain the original target population and note the limitations of extrapolated results and likelihood of remnant bias. It is also important to acknowledge that a different study may need to be conducted.

\section{Generalizability and transportability methods for estimating population average treatment effects}

Following the application of the methods in the previous sections, including assessing the plausibility of relevant assumptions, an analytic method is typically needed to generalize or transport results from randomized or observational data to a target population. These approaches have many parallels to those used to address internal validity bias. We revisit  weighting and matching-based methods and outcome regressions in  depth while additionally examining techniques that use both propensity and outcome regressions (these are often doubly  robust). To mitigate external validity bias, generalizability and transportability methods address differences in the distribution of effect modifiers between study and target populations. To do so, for weighting and matching-based approaches, these methods account for the probability of selection into the study, rather than the probability of treatment assignment. Outcome regressions require that treatment effect is allowed to vary across all effect modifiers in addition to all confounders being correctly included in the regression. 
 
Most generalizability and transportability methods have been developed for randomized data. 
When outcome data are available from both randomized studies and an observational study representative of the target population, their combination has the potential to overcome sensitivity to positivity violations for selection into the study (an issue that RCT data commonly face) as well as to unmeasured confounding (which may  afflict observational studies). Incorporating observational data in a principled manner can also shrink mean squared error. However, many such approaches do not leverage the internal validity of RCT data. The following sections will highlight some exceptions. While most approaches require individual-level study and target sample data, the Appendix highlights approaches that only use summary-level data for either the study or target sample.

\subsection{Weighting and matching methods}\label{weighting_matching}

Methods that adjust for differing baseline covariate distributions between study and target samples via weighting or matching are particularly effective when effect modifiers strongly predict selection into the study. While including unnecessary covariates can decrease precision, increase the chance of extreme weights and difficult-to-match subjects, and provide no bias reduction \citep{nie2013}, failing to include an effect modifier is typically of greater concern than including unnecessary covariates \citep{stuart2010,dahabreh2018}. 
Matching and reweighting methods 
strongly rely on common covariate support between study and target populations and perform poorly when a portion of the target population is not well-represented in the study sample or when empirical positivity violations occur. Investigators should use the estimation approach that leads to the best effect modifier balance for their study \citep{stuart2010} and strive for fewer assumptions.

\subsubsection{Matching}
Full matching and fine balance of covariate first moments (i.e., expected values) have been used in the generalizability context \citep{stuart2011,bennett2020}. \cite{stuart2011} fully match study and target sample individuals based on their propensity scores to form sets so that each matched set has at least one study and target individual. Individuals' outcomes are then reweighting by the number of target sample individuals in their matched set. This approach relies heavily on the distance metric used, which can be misled by covariates that don't affect the outcome. Fine balance of covariate first moments is a nonparametric approach for larger data that can also be used with multi-valued treatments \citep{bennett2020}. This approach matches samples to a target population to achieve fine balance on the first moments of all covariates   rather than working with the propensity score. 

Some implementations of these methods only match a subset of study individuals (hence show areas of the covariate distribution without common support), while others ensure all study and target sample individuals are matched. Matching methods require calibration for bias-variance tradeoff such as via a caliper or by choosing the ratio of study to target individuals to match. A variety of distance metrics exist; however, none specifically target effect modifiers. With unrepresentative observational data, treatment groups can first be matched based on confounding variables before matching study pairs to the target sample based on effect modifiers, or each treatment group can be separately matched to the target sample \citep{bennett2020}.

\subsubsection{Weighting}
\paragraph{Post-stratification.}
In a low-dimensional  setting with  categorical or binary covariates, one can use nonparametric post-stratification (also known as direct adjustment or subclassification), as has been done in the literature with randomized data \citep{miettinen1972,prentice2005} and with observational data in the context of instrumental variables \citep{angrist2013}. Post-stratification consists of obtaining estimates for each stratum of effect modifiers, then reweighting these estimates to reflect the effect modifier distribution in the target population, i.e., $\hat{E}(Y^a) = 1/n \sum_{l=1}^L n_{l} \bar{Y}^a_{l}$, where $L$ is the number of strata, $n_{l}$ is the target sample size in stratum $l$, $n=\sum_{l=1}^L n_{l}$, and $\bar{Y}^a_{l}$ is an estimate from study sample data of the potential outcome on treatment $a$ in stratum $l$, commonly the stratum-specific sample mean for subjects on treatment $a$ \citep{miettinen1972,prentice2005}. 

Post-stratification only requires stratum-specific summary data and closed-form variance formulas are often available. However, empty strata quickly become an issue when dealing with continuous variables or many stratifying variables. Conversely, if insufficient strata are used, residual external validity bias will remain, which is particularly problematic in small samples \citep{tipton2017a}. 
To combat this, inference can be pooled across strata using multilevel regression with post-stratification \citep{pool1964,gelman1997,park2004,kennedy2019}. 

For higher dimensional settings or with continuous covariates, more flexible nonparametric approaches can be applied, such as maximum entropy weighting, where study strata are reweighted to the distribution in the target sample \citep{hartman2015}. When target and study populations differ on post-treatment variables such as adherence, principal stratification can be used to estimate PATEs by classifying subjects into never-taker, always-taker, and complier categories \citep{frangakis2009}.

\paragraph{Estimating using the propensity for study selection.}
Most weighting approaches use a propensity of selection regression to construct weights. They rely on correct specification of the propensity score regression and sufficient overlap in propensity scores between study subjects and target sample individuals not in the study. These approaches have the additional advantage of allowing one set of weights to be used for treatment effects related to multiple outcomes. The most straightforward weighting approaches tend to have large variances in the presence of extreme weights, give disproportionate weight to outlier observations, and produce outcome estimates outside the support of the outcome variable. Weight standardization can address these issues, as can weight trimming, although the latter induces bias by changing the target population of interest, hence requiring a careful bias-variance trade-off. 

 Inverse probability of participation weighting (IPPW), a Horvitz-Thompson-like  approach \citep{horvitz1952}, is the most common weighting technique for generalizability  \citep{flores2013,baker2013,lesko2017,westreich2017,correa2018,dahabreh2018,dahabreh2019a}. Most simply, IPPW weights the outcome for each study individual on treatment \textit{a} by the inverse probability (propensity) of being in the study. Weights have been developed for estimating PATEs, including those that incorporate treatment assignment to account for covariate imbalances in an RCT or for confounding in an observational study. The observed outcomes are reweighted to obtain the potential outcomes for each treatment group $a$: $E(Y^a) = \frac{1}{n}\sum_{i=1}^{n} w_i Y_i$ with

\resizebox{0.95\hsize}{!}{
    \begin{minipage}{\linewidth}
    \begin{align*}
        w_i &= \frac{1}{\pi_{s,i}}I(S_i=1) I(A_i=a) &\mbox{ for random treatment assignment \citep{lesko2017}} \\
        w_i &= \frac{1}{\pi_{s,i} \pi_{a,i}} I(S_i=1) I(A_i=a) &\mbox{ more generally \citep{stuart2011,dahabreh2019a}}, 
    \end{align*}
    \end{minipage}
}

\noindent where $I(S_i=1)$ is the indicator for being in the study, $I(A_i=a)$ is the indicator for being assigned treatment $a$, $\pi_{s,i}=P(S_i=1|X_i)$ is the propensity score for selection into the study and $\pi_{a,i}=P(A_i=a|S_i=1,X_i)$ is the propensity score for assignment to treatment $a$ in the study. 

Individual-level data are typically required, although one can also use joint covariate distributions from group-level data \citep{cole2010} or univariate moments (e.g., means, variances) with additional assumptions \citep{signorovitch2010,phillippo2018}. Because IPPW only uses study individuals on a given treatment to estimate potential outcomes for that treatment, power can become an issue, particularly for multi-level treatments. These methods also perform poorly when study selection probabilities are small, which can be a common occurrence for generalizability \citep{tipton2013}. IPPW weights have also been developed for regression parameters in a generalized linear model \citep{haneuse2009}, as well as for Cox model hazard ratios and baseline risks \citep{cole2010,pan2008}.

For transportability to the target population $S=0$, odds of participation weights are used rather than inverse probability of participation weights \citep{westreich2017,dahabreh2018}. This corresponds to the estimator $E(Y^a|S=0) = \frac{1}{n}\sum_{i=1}^{N} w_i Y_i$ with $N=n+n_{s}$ and weights \citep{dahabreh2018}: 

\resizebox{1\hsize}{!}{
    \begin{minipage}{\linewidth}
    \begin{align*}
    w_i &= \frac{1-\pi_{s,i}}{\pi_{s,i}\pi_{a,i}}I(S_i=1) I(A_i=a). \\
    \end{align*}
    \end{minipage}
}

\noindent To address potentially unbounded outcome estimates, standardization then replaces $n$ by the sum of the weights, which normalizes the weights to sum to 1 \citep{dahabreh2018,dahabreh2019a}. The resulting estimator will be more stable, bounded by the range of the observed outcomes, and perform better when the target sample is much larger than the study.

Under regularity conditions, estimates derived using IPPW are consistent and asymptotically normal 
\citep{lunceford2004,pan2008,cole2010,correa2018,buchanan2018}. Variance for the IPPW estimator can be obtained through either a bootstrap approach or robust sandwich estimators. The latter may be difficult to calculate \citep{haneuse2009} and  bootstrap methods for IPPW have been shown to perform better when there is substantial treatment effect heterogeneity or smaller sample sizes \citep{chen2017a,tipton2017a}.

Propensity scores can also be used in the context of post-stratification,  weighting or matching individuals within strata. RCT individuals are divided into strata defined by their propensity scores; quintiles are commonly used, based on results showing that this approach may remove over 90\% of bias \citep{omuircheartaigh2014}. Effects are  estimated using sample data within each subgroup, such as through separate regressions or a joint parametric regression with fixed effects for subgroups and interaction terms for subgroups by RCT status. Results can   then be reweighted based on the number of target sample individuals in each subgroup 
\citep{omuircheartaigh2014}. Alternatively,  the target sample can be matched to RCT individuals within the same propensity score stratum \citep{tipton2013}. 

The post-stratification estimator is asymptotically normal and closed-form variance estimates exist for independent strata \citep{omuircheartaigh2014,lunceford2004}. Compared to IPPW, strata reweighting is more  likely to be  numerically stable and easily implementable when treatment assignment is done at the group level (e.g., cluster-randomized trials). However, stratification implicitly assumes that treatment effects are identical for study and target patients in the same stratum; this assumption is rarely met, resulting in residual confounding and inconsistent estimates \citep{lunceford2004}. It also relies on the assumptions of treatment effect heterogeneity being fully captured by the propensity score for treatment and that outcomes are continuous and bounded. With too few strata, bias reduction will be insufficient; conversely, too many strata can lead to small strata counts and unstable estimates \citep{stuart2010,tipton2017a}.

Propensity strata approaches have also been used to address positivity of treatment assignment violations within the target sample in the setting where outcome data are available from both a randomized and observational study \citep{rosenman2018}. \cite{rosenman2020} present an extension which aims to adjust for potential unmeasured confounding bias.

\subsection{Outcome regression  methods}
\subsubsection{Outcome data from one study.  }

 Outcome regressions, also known as response surface modeling, have not been as extensively developed for generalizability and transportability compared to propensity-based approaches. Broadly speaking, outcome regressions approaches fit an outcome regression in study sample data to estimate conditional means, then obtain PATEs by marginalizing over (i.e., standardizing to) the target sample covariate distribution by predicting counterfactuals for the target sample: $\hat{E}(Y^a) = \frac{1}{n}\sum_{i=1}^{n} \hat{E}(Y_i | S_i=1, A_i=a, X_i)$. If the target sample is not a simple random sample from the target population, this would be a weighted average using sampling weights \citep{kim2018}. 
 
 Outcome regression approaches are particularly effective when effect modifiers strongly predict the outcome and when the outcome is common but selection into the study is rare. They are also convenient for exploring PCATEs. These approaches can yield better precision than weighting or matching-based methods because they can adjust both for confounders, effect-modifiers, and factors only predictive of the outcome, thus decreasing variance in the estimate. They are simple to implement when an outcome regression for confounding adjustment has already been fit and accounts for all relevant effect modifiers. The same regression that was used to estimate impacts within the study can then be used to predict counterfactuals in the target sample. Outcome regression methods can be used with either randomized or observational study data, but have been used most frequently in RCTs. In the presence of significant non-overlap between the target and study samples, outcome regressions rely on heavy extrapolation \citep{kern2016,attanasio2003}, often with no corresponding inflation of the variance to reflect uncertainty in the resulting estimates.

The simplest approach is an ordinary least squares outcome regression \citep{flores2013,kern2016,elliott2017,dahabreh2018,dahabreh2019a}. An outcome regression is fit with interaction terms between treatment and all effect modifiers before predicting  counterfactual outcomes for the target sample (the marginalization step).
\cite{dahabreh2018} showed the consistency of this type of outcome regression for the PATE. For RCTs, separate regressions are recommended for each treatment group to better capture treatment effect heterogeneity \citep{dahabreh2019a}, although this approach precludes borrowing information across treatment groups, which is possible with machine learning methods that discover treatment effect heterogeneity.

Among these machine learning techniques is BART, which is the most commonly used data-adaptive outcome regression approach for generalizability and transportability \citep{chipman2007,chipman2010,kern2016,hill2011}. Tree-based methods, including BART, were briefly introduced in Section~\ref{TEH}. BART models the outcome as a sum of trees with linear additive terms and  a regularization prior. 
BART addresses external validity bias via its data-driven discovery of treatment effect heterogeneity and strengths of the method include its ability to obtain confidence intervals from the posterior distribution 
\citep{hill2011,green2012}. 
However, BART credible intervals show undercoverage when the target population differs substantially from the RCT \citep{hill2011}.

Data availability may challenge these outcome regression approaches. When the covariates in the target sample aren't available in the study sample, or vice versa, but the SATE can be expected to be approximately unbiased for the PATE, the SATE estimates' credible intervals can be expanded to account for uncertainty in the target population covariate distribution \citep{hill2011}. 

\subsubsection{Outcome data from multiple studies.}
Here, we consider meta-analytic approaches for summary-level data as well as studies that combine individual-level  data from more than one study (for example, one randomized and one observational study). Much of the literature has focused on meta-analytic techniques using summary-level study data and no target sample covariate information. This body of bias-adjusted meta-analysis methods largely does not explicitly define a target population for whom inference is desired, but rather relies on subjective investigator judgments of the levels of bias in each study, specified using bias functions or priors in a Bayesian framework. 
\cite{eddy1989} presents the first such approach, the confidence profile method for combining chains of evidence. Likelihoods are adjusted for different study designs' (investigator-specified) internal and external validity biases; uncertainty around these biases are incorporated through prior distributions. Various subsequent Bayesian hierarchical models have been developed, such as a 3-level model \citep{prevost2000} with the levels corresponding to models of the observed evidence, variability between studies, and variability between study types (randomized vs. observational). When available, covariate information can be added to the models to address effect heterogeneity. 
Effectively, this estimator averages across the internal and external validity biases of the studies and therefore is only unbiased when the external validity bias in the RCT exactly `cancels' the internal validity bias in the observational data \citep{kaizar2011}. 

Other meta-analysis studies leveraging summary-level data separately specify internal and external validity bias parameters for an explicit target population and down-weight studies with higher risk of bias. One such example is the bias adjusted meta-analysis approach by \cite{turner2009}, which 
presents a checklist  that subjectively quantifies the extent of internal and external validity bias for each study and then weighs studies' average outcomes by the extent of bias. 
\cite{greenland2005}  pool across observational case-control studies using a Bayesian meta-sensitivity model with bias parameters to separately permit consideration of misclassification, non-response, and unmeasured confounding. In the intermediate setting where individual-level data is available in the study but only covariate moments (e.g., means, variances) are available in the target setting, \cite{phillippo2018} present an outcome regression approach for indirect treatment comparison across RCTs.

When individual-level outcome data is available in the target sample or from multiple studies, data can be combined into one joint dataset for outcome regression analysis if the outcome regression can be expected to be the same across studies \citep{kern2016}. Such an approach can be preferential to IPPW, which uses only study and not target sample outcome data \citep{kern2016}. However, it will be dominated by observational data results (and their potential biases) when observational subjects constitute the majority of the joint dataset, effectively result in a weighted average across studies, weighted by the proportion of subjects in each study.   

Hierarchical Bayesian evidence synthesis is the only outcome regression approach we identified that attempts to empirically adjust for unobserved confounding when estimating effects for observational patients who are not well-represented in the RCTs \citep{verde2016,verde2019}. 
Summary-level RCT data are combined with individual-level observational data through a weighting approach in which the control group event rate is assumed to be similar across all studies and a study quality bias term is added to the observational studies' outcome regression to account for unmeasured confounding or other uncontrolled biases and to inflate variance.  Alternatively, \cite{gechter2015} derive bounds on the PATE and PCATE when transporting from an RCT to a target sample with outcome data (all untreated).

\subsection{Combined propensity score and outcome regression methods}
\subsubsection{Outcome data from one study.}

Double robust methods for generalizability and transportability typically combine outcome and propensity of selection regressions. They are asymptotically unbiased when at least one of these regression functions is consistently estimated, and if both are consistently estimated, asymptotically efficient. However, if neither  regression is estimated consistently, the mean squared error may be worse than using a propensity or outcome regression alone. Incorporation of flexible modeling approaches can help mitigate regression misspecification. 
Three asymptotically locally efficient double robust approaches have been developed in randomized data: a targeted maximum likelihood estimator (TMLE) for instrumental variables \citep{rudolph2017}, which is a semiparametric substitution estimator, the estimating equation-based augmented inverse probability of participation weighting (A-IPPW) \citep{dahabreh2018,dahabreh2019a}, and an augmented calibration weighting estimator that can also incorporate outcome information from the target sample when it is available \citep{dong2020}. 

The TMLE was developed for transportability in an encouragement design setting (i.e.,  intervention focused on encouraging individuals in the treatment group to participate in the intervention)  with instrumental variables \citep{rudolph2017} and has also been used for generalizability \citep{schmid2020}. Three different PATE estimators were developed: intent to treat, complier, and as-treated. All use an outcome regression to obtain an initial estimate, then adjust that estimate with a fluctuation function using a clever covariate $C$, which is derived from the efficient influence curve and incorporates the propensity of selection information in a bias reduction step. 
For example, for the intent to treat PATE, the fluctuation function takes the form: $\text{logit}(\hat{E}(Y|S=1,A,Z,X)+\epsilon C)$, where 
$$C = \frac{I(S=1,A=a)}{P(A=a|S=1,X)P(S=1)}\frac{P(Z=z|S=0,A=a,X)P(X|S=0)}{P(Z=z|S=1,A=a,X)P(X|S=1)}$$
\noindent and $Z$ corresponds to the intervention taken (whereas $A$ corresponds to the assigned intervention, as before).
The approach allows outcome and propensity regressions to be flexibly fit, for example, using an ensemble of machine learning algorithms. Variances are calculated from the  influence curve. 

A-IPPW has been developed both for generalizing results to estimate PATEs for all trial-eligible individuals \citep{dahabreh2019a,dahabreh2019b} and for transporting results to estimate PATEs for trial-eligible individuals not included in a trial \citep{dahabreh2018}. Three double robust estimating equation-based estimators are presented: A-IPPW, A-IPPW with normalized weights that sum to 1 to ensure bounded estimates, and a weighted outcome regression estimator  using participation weights. The non-normalized A-IPPW estimators are as follows, with $w_i$ the same as for IPPW: 

\resizebox{0.95\hsize}{!}{
    \begin{minipage}{\linewidth}
    \begin{align*}
    &\frac{1}{n}\sum_{i=1}^{n}\{w_i\{Y_i-\hat{E}(Y_i|S_i=1,A_i=a,X_i)\}+\hat{E}(Y_i|S_i=1,A_i=a,X_i)\} &\mbox{for generalizability} \\
    &\frac{1}{n}\sum_{i=1}^{N}\{w_i\{Y_i-\hat{E}(Y_i|S_i=1,A_i=a,X_i)\}+\{1-I(S_i=1)\}\hat{E}(Y_i|S_i=1,A_i=a,X_i)\} &\mbox{for transportability}.
    \end{align*}
    \end{minipage}
}

\noindent Variance can be derived using empirical sandwich estimates or using a nonparametric bootstrap. As these estimators are partial M-estimators, they can produce estimates outside bounds if the outcome regression is not well-chosen and they may have multiple solutions. 

Several other double robust estimators for transportability resemble the IPPW estimator, with sampling weights derived through alternative approaches that do not rely on propensity scores \citep{josey2020,josey2020a,dong2020}. For example, the semiparametric and efficient augmented weighting estimator by \cite{dong2020} calibrates the RCT covariate distribution to match that of the sampling-weighted target sample.

An alternative reweighted outcome regression method for observational data does not claim double robustness and draws from the unsupervised domain adaptation literature. In general, unsupervised domain adaptation methods aim to make predictions for a target sample (the ``target domain'') when outcomes are only observed in the study sample (``source domain''). The approach of \cite{johansson2018} is a regularized neural network estimator for PCATE parameters that jointly learns representations from the data and a reweighting function. Representational learning creates balance between the study and target covariate distributions and between treated and control distributions in a representational space so that predictors use information common across these distributions and focus on covariates predictive of the outcome. In this learned representational space, results are then re-weighted to minimize an upper bound on the expected value of the loss function under the target covariate distribution.  Propensity scores can also be used to reweight a likelihood function, as done by \cite{nie2013} in an RCT setting for calibrating  control outcomes from prior studies to the trial target sample. Similarly, \cite{flores2013} reweight an outcome regression to the target sample.

 \subsubsection{Outcome data from multiple studies.}

Several methods have combined randomized and observational data sources such that that they retain the internal validity of the randomized data and the external validity of the target sample observational data. These approaches broadly rely on the assumption that the relationship between unmeasured confounders and potential outcomes is the same in the RCT as in the target sample, which is a weaker assumption than that of no unmeasured confounding required by most of the methods described thus far. One study combined individual-level data from several RCTs to transport results to the target sample, extending the A-IPPW estimator (as well as corresponding IPPW and outcome regression estimators) to the multi-study setting \citep{dahabreh2019}. The remainder of the section discusses approaches that combine randomized and observational data.

When differences in effect modifiers between the RCT and target population are known (e.g., by inclusion and exclusion criteria), cross-design synthesis meta-analysis is a method for combining randomized and observational study data while capitalizing on the internal validity of the randomized data and the external validity of the observational data \citep{begg1992,greenhouse2017}. It provides a means for estimating treatment effects for patients excluded from the RCT and can use summary-level RCT data if outcomes are available by relevant patient subgroups, although can only accommodate a limited number of strata of relevant effect modifiers. 

Cross-design synthesis meta-analysis effectively assumes a constant amount of unmeasured confounding across patients eligible and ineligible for the RCTs \citep{kaizar2011}. This approach will have smaller bias than use of randomized or observational data alone under various common data scenarios and, across simulations, shows better coverage through smaller bias and increased variance \citep{kaizar2011}.

When differences between RCT and target populations are less well understood, there are continuous effect modifiers, or a higher dimensional set of effect modifiers, one can use Bayesian calibrated risk-adjusted regressions \citep{varadhan2016,henderson2017}. This parametric approach requires individual-level information from observational and randomized studies, leveraging outcome regressions and calibration using  the propensity of selection. The target population is assumed to be represented by a subset of the observational data; the RCT data are likewise assumed to be represented by a (potentially different) subset of the observational data. 
The calibrated risk-adjusted model performs well when there is poor overlap between RCT and target data; however, it relies on the observational dataset having substantial effect modifier overlap with both the target sample and RCT. 
Robust variance formulas or bootstrapping can be used to obtain confidence intervals.

A 2-step frequentist approach for consistently estimating PCATE parameters has been developed to estimate effects in a target population represented by observational data \citep{kallus2018}. It begins with outcome regressions for each treatment group of the observational data, or a flexible regression that captures effect heterogeneity. Observational data are then standardized to the RCT population before `debiasing' their estimates using RCT data by including a correction term that can depend on measured covariates. This method relies on the assumption that calibrating internal validity bias in the subset of the observational data distribution overlapping with RCT data appropriately calibrates the bias for the entire target sample. The 2-step approach would therefore not necessarily decrease bias if the covariate distribution is highly imbalanced, resulting in average biases that are quite different between the RCT overlapping vs. nonoverlapping subsets of the target sample.

\cite{lu2019} present an approach that, unlike the above methods, assumes no unmeasured confounding in the observational data when combining RCT and comprehensive cohort study data (where patients who decline randomization are enrolled in a parallel observational study). They use semiparametric double robust estimators that can incorporate flexible regressions.

\section{Discussion}
Obtaining unbiased estimates for a relevant target population requires applying generalizability or transportability methods in studies that meet required identifiability assumptions. The internal validity of randomized trials is not sufficient to obtain unbiased causal effects; external validity also needs to be considered.  In this synthesis, we have discussed (1) sources of external validity bias and study designs to address it, (2) defining an estimand in a target population of interest, (3) the identifiability assumptions underpinning generalizability and transportability approaches, (4) a variety of approaches for quantifying the relevant dissimilarity between study and target samples and assessing treatment effect heterogeneity, and (5) a variety of matching and weighting methods, outcome regression approaches, and techniques that use both outcome and propensity regressions that generalize or transport from randomized and observational studies to a target population.   These approaches have been applied across diverse settings from RCT results transported to patients represented in registries to cluster-randomized educational intervention trials generalized to broader geographic areas. Across a variety of settings, it is important to estimate results for populations that go beyond the study population. We suggest the following considerations for researchers.

\textit{ Make efforts to explicitly define the target population(s) and identify the study population from which your study sample data is a simple random sample.} Describing the study population may be a difficult task, and there may not be a practically meaningful population that is representative of your study sample data. However, this clarity will allow you to compare and, when feasible, better-align the study sample data to the target population. Discussion regarding target population(s) should be guided by the ensuing decisions the study aims to inform as well as practical considerations (e.g., lack of certain subgroups in your study). These considerations may require iteration between feasibility and the desired study aims as well as careful discussion amidst study collaborators. When combining across studies, meta-analyses should likewise carefully specify target population(s) for inference and incorporate considerations of treatment effect heterogeneity or demonstrate that effect heterogeneity is not a concern. Without transparency in the target population(s), a study cannot estimate well-defined treatment effects nor can readers judge the generalizability of study results to any other population of interest. 

 \textit{  Plan for generalization in your study design, when feasible, including writing generalizability considerations into your grant or study objectives.} Enroll randomized study participants or design observational study inclusion and exclusion criteria to have the study sample be representative of the target population, or fully capturing the heterogeneity of effect modifiers. Collect data on likely treatment effect modifiers that are associated with study participation. Attempt to identify and mitigate potential sources of missingness or selection bias. If possible, collect baseline characteristics and outcome data on study nonparticipants who are part of the target population. Otherwise, identify external sources of data that might inform the composition of your target population with respect to effect modifiers and work towards aligning variables between these target sample data sources and your study.

   \textit{ Clearly describe the internal and external validity assumptions needed to identify the treatment effect as they relate to your study.} Substantively assess the justifiability of these internal and external validity assumptions. To the extent possible, test the validity of the assumptions and perform sensitivity analyses to assess the impact of assumption violations. 

  \textit{ Quantify the dissimilarity between the study and target populations using at least one method.} Ideally, use multiple methods, as they each tell different parts of the story: examine univariate and joint distributions of effect modifiers, differences in the propensity to participate in the study, and (if outcome information is available in the target sample) differences in outcomes between study and target subjects on the same treatment. If differences are identified, one should investigate which subpopulations drive those differences and assess whether they have heterogeneous treatment effects. In addition to examining subject characteristics, assess whether differences exist in the setting, treatment, or outcome. 

   \textit{ To obtain causal estimates when the target and study populations differ with respect to effect modifiers, incorporate at least one generalizability or transportability estimator.} Alternatively, at the minimum, assess and describe sources of effect heterogeneity and whether they're likely to differ for the target population. Derive estimates using as much data as possible (e.g., when outcome data is available, use it in a principled way). The choice of method for external validity bias adjustment may be restricted by data availability (e.g., summary-level vs. individual-level data) but should be driven by similar principles as those that guide the choice between outcome regressions, matching and weighting methods, and double robust approaches for confounding adjustment \citep{vanderlaan2003,neugebauer2005,vanderlaan2011}. Flexible nonparametric and semiparametric models and estimators that  use ensemble machine learning minimize the need for strict parametric assumptions and have the potential to perform the best \citep{kern2016}. 

  \textit{ For both methods developers and applied researchers, we recommend releasing publicly available  code alongside the paper and providing details for implementation.} Published code facilitates replicability and accessibility of  methods for future research and applied use. A substantial barrier to the adoption of new statistical methods, including advances in generalizability and transportability, is the lack of available computational tools.

While much of the causal inference literature has focused on issues of internal validity, both internal and external validity are necessary for valid inference. When treatment effect heterogeneity exists, as is often the case, study results may not hold for a target population of interest. Approaches to address internal validity biases can be borrowed  to improve upon methods for addressing external validity bias. This review presents a framework for such analysis and summarizes different choices for estimators that can be used to generalize or transport results to a population different from the one under study. It brings together diverse cross-disciplinary literature to provide guidance both for applied and methods researchers. Improving the incorporation of results from observational studies, including electronic health databases, can lead to better inference for policy-relevant populations with reduced bias and improved precision.

\section{Acknowledgments}

This research was supported by NIH New Innovator Award DP2MD012722 and NIH training grants T32LM012411 and T32ES07142. The authors thank Sebastien Haneuse, Francesca Dominici, and Laura Hatfield for helpful feedback on this work as well as seminar and conference audiences at the Harvard Program on Causal Inference, Mathematica Policy Research, Harvard--Stanford Health Policy Data Science Lab, 2018 Harvard Data Science Initiative Conference, and 2020 NIA Workshop on Applications of Machine Learning to Improve Healthcare Delivery for Older Adults.

\bibliographystyle{imsart-nameyear}
\bibliography{Generalizability_Review_Library_month.bib}

\begin{thebibliography}{144}

\bibitem[\protect\citeauthoryear{Ackerman et~al.}{2019}]{ackerman2019}
\begin{barticle}[author]
\bauthor{\bsnm{Ackerman},~\bfnm{Benjamin}\binits{B.}},
  \bauthor{\bsnm{Schmid},~\bfnm{Ian}\binits{I.}},
  \bauthor{\bsnm{Rudolph},~\bfnm{Kara~E}\binits{K.~E.}},
  \bauthor{\bsnm{Seamans},~\bfnm{Marissa~J}\binits{M.~J.}},
  \bauthor{\bsnm{Susukida},~\bfnm{Ryoko}\binits{R.}},
  \bauthor{\bsnm{Mojtabai},~\bfnm{Ramin}\binits{R.}} \AND
  \bauthor{\bsnm{Stuart},~\bfnm{Elizabeth~A}\binits{E.~A.}}
(\byear{2019}).
\btitle{Implementing Statistical Methods for Generalizing Randomized Trial
  Findings to a Target Population}.
\bjournal{Addictive behaviors}
\bvolume{94}
\bpages{124--132}.
\end{barticle}
\endbibitem

\bibitem[\protect\citeauthoryear{Allcott}{2015}]{allcott2015}
\begin{barticle}[author]
\bauthor{\bsnm{Allcott},~\bfnm{Hunt}\binits{H.}}
(\byear{2015}).
\btitle{Site Selection Bias in Program Evaluation}.
\bjournal{The Quarterly journal of economics}
\bvolume{130}
\bpages{1117--1166}.
\end{barticle}
\endbibitem

\bibitem[\protect\citeauthoryear{Allcott and Mullainathan}{2012}]{allcott2012}
\begin{barticle}[author]
\bauthor{\bsnm{Allcott},~\bfnm{Hunt}\binits{H.}} \AND
  \bauthor{\bsnm{Mullainathan},~\bfnm{Sendhil}\binits{S.}}
(\byear{2012}).
\btitle{External Validity and Partner Selection Bias}.
\bjournal{National Bureau of Economic Research Working Paper Series}
\bvolume{18373}
\bpages{53}.
\end{barticle}
\endbibitem

\bibitem[\protect\citeauthoryear{Andrews and Oster}{2017}]{andrews2017}
\begin{btechreport}[author]
\bauthor{\bsnm{Andrews},~\bfnm{Isaiah}\binits{I.}} \AND
  \bauthor{\bsnm{Oster},~\bfnm{Emily}\binits{E.}}
(\byear{2017}).
\btitle{Weighting for External Validity}
\btype{Technical Report} No. \bnumber{w23826},
\bpublisher{{National Bureau of Economic Research}},
\baddress{{Cambridge, MA}}.
\bdoi{10.3386/w23826}
\end{btechreport}
\endbibitem

\bibitem[\protect\citeauthoryear{Angrist and
  {Fern{\'a}ndez-Val}}{2013}]{angrist2013}
\begin{bincollection}[author]
\bauthor{\bsnm{Angrist},~\bfnm{Joshua~D}\binits{J.~D.}} \AND
  \bauthor{\bsnm{{Fern{\'a}ndez-Val}},~\bfnm{Iv{\'a}n}\binits{I.}}
(\byear{2013}).
\btitle{{{ExtrapoLATE}}-Ing: External Validity and Overidentification in the
  {{LATE}} Framework}.
In \bbooktitle{Advances in {{Economics}} and {{Econometrics}}}
\bpages{401--434}.
\bpublisher{{Cambridge University Press}}.
\end{bincollection}
\endbibitem

\bibitem[\protect\citeauthoryear{Athey and Imbens}{2016}]{athey2016}
\begin{barticle}[author]
\bauthor{\bsnm{Athey},~\bfnm{Susan}\binits{S.}} \AND
  \bauthor{\bsnm{Imbens},~\bfnm{Guido}\binits{G.}}
(\byear{2016}).
\btitle{Recursive Partitioning for Heterogeneous Causal Effects}.
\bjournal{Proceedings of the National Academy of Sciences}
\bvolume{113}
\bpages{7353--7360}.
\bdoi{10.1073/pnas.1510489113}
\end{barticle}
\endbibitem

\bibitem[\protect\citeauthoryear{Attanasio, Meghir and
  Szekely}{2003}]{attanasio2003}
\begin{barticle}[author]
\bauthor{\bsnm{Attanasio},~\bfnm{Orazio}\binits{O.}},
  \bauthor{\bsnm{Meghir},~\bfnm{Costas}\binits{C.}} \AND
  \bauthor{\bsnm{Szekely},~\bfnm{Miguel}\binits{M.}}
(\byear{2003}).
\btitle{Using Randomised Experiments and Structural Models for 'Scaling up':
  Evidence from the {{PROGRESA}} Evaluation}.
\bjournal{IFS Working Paper}
\bvolume{EWP03/05}.
\end{barticle}
\endbibitem

\bibitem[\protect\citeauthoryear{Baker et~al.}{2013}]{baker2013}
\begin{barticle}[author]
\bauthor{\bsnm{Baker},~\bfnm{Reg}\binits{R.}},
  \bauthor{\bsnm{Brick},~\bfnm{J.~Michael}\binits{J.~M.}},
  \bauthor{\bsnm{Gotway~Crawford},~\bfnm{Carol~A}\binits{C.~A.}},
  \bauthor{\bsnm{Terhanian},~\bfnm{George}\binits{G.}},
  \bauthor{\bsnm{Langer},~\bfnm{Gary}\binits{G.}},
  \bauthor{\bsnm{Bates},~\bfnm{Nancy~A}\binits{N.~A.}},
  \bauthor{\bsnm{Battaglia},~\bfnm{Mike}\binits{M.}},
  \bauthor{\bsnm{Couper},~\bfnm{Mick~P}\binits{M.~P.}},
  \bauthor{\bsnm{Dever},~\bfnm{Jill~A}\binits{J.~A.}},
  \bauthor{\bsnm{Gile},~\bfnm{Krista~J}\binits{K.~J.}},
  \bauthor{\bsnm{Tourangeau},~\bfnm{Roger}\binits{R.}},
  \bauthor{\bsnm{Valliant},~\bfnm{Richard}\binits{R.}} \AND
  \bauthor{\bsnm{Rivers},~\bfnm{Douglas}\binits{D.}}
(\byear{2013}).
\btitle{Summary Report of the Aapor Task Force on Non-Probability Sampling}.
\bjournal{Journal of survey statistics and methodology}
\bvolume{1}
\bpages{90--136}.
\end{barticle}
\endbibitem

\bibitem[\protect\citeauthoryear{Bareinboim and Pearl}{2014}]{bareinboim2014a}
\begin{bincollection}[author]
\bauthor{\bsnm{Bareinboim},~\bfnm{Elias}\binits{E.}} \AND
  \bauthor{\bsnm{Pearl},~\bfnm{Judea}\binits{J.}}
(\byear{2014}).
\btitle{Transportability from Multiple Environments with Limited Experiments:
  Completeness Results}.
In \bbooktitle{Advances in {{Neural Information Processing Systems}} 27}
(\beditor{\bfnm{Z.}\binits{Z.}~\bsnm{Ghahramani}},
  \beditor{\bfnm{M.}\binits{M.}~\bsnm{Welling}},
  \beditor{\bfnm{C.}\binits{C.}~\bsnm{Cortes}},
  \beditor{\bfnm{N.~D.}\binits{N.~D.}~\bsnm{Lawrence}} \AND
  \beditor{\bfnm{K.~Q.}\binits{K.~Q.}~\bsnm{Weinberger}}, eds.)
\bpages{280--288}.
\bpublisher{{Curran Associates, Inc.}}
\end{bincollection}
\endbibitem

\bibitem[\protect\citeauthoryear{Bareinboim and Pearl}{2016}]{bareinboim2016}
\begin{barticle}[author]
\bauthor{\bsnm{Bareinboim},~\bfnm{Elias}\binits{E.}} \AND
  \bauthor{\bsnm{Pearl},~\bfnm{Judea}\binits{J.}}
(\byear{2016}).
\btitle{Causal Inference and the Data-Fusion Problem}.
\bjournal{Proceedings of the National Academy of Sciences}
\bvolume{113}
\bpages{7345--7352}.
\bdoi{10.1073/pnas.1510507113}
\end{barticle}
\endbibitem

\bibitem[\protect\citeauthoryear{Bareinboim, Tian and
  Pearl}{2014}]{bareinboim2014}
\begin{binproceedings}[author]
\bauthor{\bsnm{Bareinboim},~\bfnm{Elias}\binits{E.}},
  \bauthor{\bsnm{Tian},~\bfnm{Jin}\binits{J.}} \AND
  \bauthor{\bsnm{Pearl},~\bfnm{Judea}\binits{J.}}
(\byear{2014}).
\btitle{Recovering from Selection Bias in Causal and Statistical Inference}.
In \bbooktitle{Proceedings of the {{Twenty}}-{{Eighth AAAI Conference}} on
  {{Artificial Intelligence}}}.
\bseries{{{AAAI}}'14}
\bpages{2410--2416}.
\bpublisher{{AAAI Press}}.
\end{binproceedings}
\endbibitem

\bibitem[\protect\citeauthoryear{Bareinboim and Tian}{2015}]{bareinboim2015}
\begin{binproceedings}[author]
\bauthor{\bsnm{Bareinboim},~\bfnm{Elias}\binits{E.}} \AND
  \bauthor{\bsnm{Tian},~\bfnm{Jin}\binits{J.}}
(\byear{2015}).
\btitle{Recovering Causal Effects from Selection Bias}.
In \bbooktitle{Proceedings of the {{Twenty}}-{{Ninth AAAI Conference}} on
  {{Artificial Intelligence}}}.
\bseries{{{AAAI}}'15}
\bpages{3475--3481}.
\bpublisher{{AAAI Press}}.
\end{binproceedings}
\endbibitem

\bibitem[\protect\citeauthoryear{Begg}{1992}]{begg1992}
\begin{barticle}[author]
\bauthor{\bsnm{Begg},~\bfnm{Colin~B.}\binits{C.~B.}}
(\byear{1992}).
\btitle{Cross Design Synthesis: A New Strategy for Medical Effectiveness
  Research. United States General Accounting Office, ({{GA0}}/{{PEMD}}-92-18)}.
\bjournal{Statistics in Medicine}
\bvolume{11}
\bpages{1627--1628}.
\end{barticle}
\endbibitem

\bibitem[\protect\citeauthoryear{Bell et~al.}{2016}]{bell2016}
\begin{barticle}[author]
\bauthor{\bsnm{Bell},~\bfnm{Stephen~H.}\binits{S.~H.}},
  \bauthor{\bsnm{Olsen},~\bfnm{Robert~B.}\binits{R.~B.}},
  \bauthor{\bsnm{Orr},~\bfnm{Larry~L.}\binits{L.~L.}} \AND
  \bauthor{\bsnm{Stuart},~\bfnm{Elizabeth~A.}\binits{E.~A.}}
(\byear{2016}).
\btitle{Estimates of External Validity Bias When Impact Evaluations Select
  Sites Nonrandomly}.
\bjournal{Educational Evaluation and Policy Analysis}
\bvolume{38}
\bpages{318--335}.
\bdoi{10.3102/0162373715617549}
\end{barticle}
\endbibitem

\bibitem[\protect\citeauthoryear{Benchimol et~al.}{2015}]{benchimol2015}
\begin{barticle}[author]
\bauthor{\bsnm{Benchimol},~\bfnm{Eric~I.}\binits{E.~I.}},
  \bauthor{\bsnm{Smeeth},~\bfnm{Liam}\binits{L.}},
  \bauthor{\bsnm{Guttmann},~\bfnm{Astrid}\binits{A.}},
  \bauthor{\bsnm{Harron},~\bfnm{Katie}\binits{K.}},
  \bauthor{\bsnm{Moher},~\bfnm{David}\binits{D.}},
  \bauthor{\bsnm{Petersen},~\bfnm{Irene}\binits{I.}},
  \bauthor{\bsnm{S{\o}rensen},~\bfnm{Henrik~T.}\binits{H.~T.}},
  \bauthor{\bsnm{{von Elm}},~\bfnm{Erik}\binits{E.}},
  \bauthor{\bsnm{Langan},~\bfnm{Sin{\'e}ad~M.}\binits{S.~M.}} \AND
  \bauthor{\bsnm{{RECORD Working Committee}}}
(\byear{2015}).
\btitle{The {{REporting}} of Studies Conducted Using Observational
  Routinely-Collected Health Data ({{RECORD}}) Statement}.
\bjournal{PLOS Medicine}
\bvolume{12}
\bpages{e1001885}.
\bdoi{10.1371/journal.pmed.1001885}
\end{barticle}
\endbibitem

\bibitem[\protect\citeauthoryear{Bennett, Vielma and
  Zubizarreta}{2020}]{bennett2020}
\begin{barticle}[author]
\bauthor{\bsnm{Bennett},~\bfnm{Magdalena}\binits{M.}},
  \bauthor{\bsnm{Vielma},~\bfnm{Juan~Pablo}\binits{J.~P.}} \AND
  \bauthor{\bsnm{Zubizarreta},~\bfnm{Jos{\'e}~R}\binits{J.~R.}}
(\byear{2020}).
\btitle{Building Representative Matched Samples with Multi-Valued Treatments in
  Large Observational Studies}.
\bjournal{Journal of computational and graphical statistics}
\bvolume{29}
\bpages{744--757}.
\end{barticle}
\endbibitem

\bibitem[\protect\citeauthoryear{Buchanan et~al.}{2018}]{buchanan2018}
\begin{barticle}[author]
\bauthor{\bsnm{Buchanan},~\bfnm{Ashley~L}\binits{A.~L.}},
  \bauthor{\bsnm{Hudgens},~\bfnm{Michael~G}\binits{M.~G.}},
  \bauthor{\bsnm{Cole},~\bfnm{Stephen~R}\binits{S.~R.}},
  \bauthor{\bsnm{Mollan},~\bfnm{Katie~R}\binits{K.~R.}},
  \bauthor{\bsnm{Sax},~\bfnm{Paul~E}\binits{P.~E.}},
  \bauthor{\bsnm{Daar},~\bfnm{Eric~S}\binits{E.~S.}},
  \bauthor{\bsnm{Adimora},~\bfnm{Adaora~A}\binits{A.~A.}},
  \bauthor{\bsnm{Eron},~\bfnm{Joseph~J}\binits{J.~J.}} \AND
  \bauthor{\bsnm{Mugavero},~\bfnm{Michael~J}\binits{M.~J.}}
(\byear{2018}).
\btitle{Generalizing Evidence from Randomized Trials Using Inverse Probability
  of Sampling Weights}.
\bjournal{Journal of the Royal Statistical Society. Series A, Statistics in
  society}
\bvolume{181}
\bpages{1193--1209}.
\end{barticle}
\endbibitem

\bibitem[\protect\citeauthoryear{Burchett, Umoquit and
  Dobrow}{2011}]{burchett2011}
\begin{barticle}[author]
\bauthor{\bsnm{Burchett},~\bfnm{Helen}\binits{H.}},
  \bauthor{\bsnm{Umoquit},~\bfnm{Muriah}\binits{M.}} \AND
  \bauthor{\bsnm{Dobrow},~\bfnm{Mark}\binits{M.}}
(\byear{2011}).
\btitle{How Do We Know When Research from One Setting Can Be Useful in Another?
  {{A}} Review of External Validity, Applicability and Transferability
  Frameworks}.
\bjournal{Journal of health services research \& policy}
\bvolume{16}
\bpages{238--244}.
\end{barticle}
\endbibitem

\bibitem[\protect\citeauthoryear{Cahan, Cahan and Cimino}{2017}]{cahan2017}
\begin{barticle}[author]
\bauthor{\bsnm{Cahan},~\bfnm{Amos}\binits{A.}},
  \bauthor{\bsnm{Cahan},~\bfnm{Sorel}\binits{S.}} \AND
  \bauthor{\bsnm{Cimino},~\bfnm{James~J.}\binits{J.~J.}}
(\byear{2017}).
\btitle{Computer-Aided Assessment of the Generalizability of Clinical Trial
  Results}.
\bjournal{International Journal of Medical Informatics}
\bvolume{99}
\bpages{60--66}.
\bdoi{10.1016/j.ijmedinf.2016.12.008}
\end{barticle}
\endbibitem

\bibitem[\protect\citeauthoryear{Chan}{2017}]{chan2017}
\begin{barticle}[author]
\bauthor{\bsnm{Chan},~\bfnm{Wendy}\binits{W.}}
(\byear{2017}).
\btitle{Partially Identified Treatment Effects for Generalizability}.
\bjournal{Journal of Research on Educational Effectiveness}
\bvolume{10}
\bpages{646--669}.
\bdoi{10.1080/19345747.2016.1273412}
\end{barticle}
\endbibitem

\bibitem[\protect\citeauthoryear{Chen and Kaizar}{2017}]{chen2017a}
\begin{barticle}[author]
\bauthor{\bsnm{Chen},~\bfnm{Ziyue}\binits{Z.}} \AND
  \bauthor{\bsnm{Kaizar},~\bfnm{Eloise}\binits{E.}}
(\byear{2017}).
\btitle{On Variance Estimation for Generalizing from a Trial to a Target
  Population}.
\bjournal{arXiv:1704.07789 [stat]}.
\end{barticle}
\endbibitem

\bibitem[\protect\citeauthoryear{Chen and Wong}{2018}]{chen2018}
\begin{bbook}[author]
\bauthor{\bsnm{Chen},~\bfnm{Caroline}\binits{C.}} \AND
  \bauthor{\bsnm{Wong},~\bfnm{Riley}\binits{R.}}
(\byear{2018}).
\btitle{Black Patients Miss out on Promising Cancer Drugs}.
\end{bbook}
\endbibitem

\bibitem[\protect\citeauthoryear{Chen et~al.}{2017}]{chen2017}
\begin{barticle}[author]
\bauthor{\bsnm{Chen},~\bfnm{Shuai}\binits{S.}},
  \bauthor{\bsnm{Tian},~\bfnm{Lu}\binits{L.}},
  \bauthor{\bsnm{Cai},~\bfnm{Tianxi}\binits{T.}} \AND
  \bauthor{\bsnm{Yu},~\bfnm{Menggang}\binits{M.}}
(\byear{2017}).
\btitle{A General Statistical Framework for Subgroup Identification and
  Comparative Treatment Scoring}.
\bjournal{Biometrics}
\bvolume{73}
\bpages{1199--1209}.
\bdoi{10.1111/biom.12676}
\end{barticle}
\endbibitem

\bibitem[\protect\citeauthoryear{Chen et~al.}{2020}]{chen2020}
\begin{barticle}[author]
\bauthor{\bsnm{Chen},~\bfnm{Irene~Y}\binits{I.~Y.}},
  \bauthor{\bsnm{Pierson},~\bfnm{Emma}\binits{E.}},
  \bauthor{\bsnm{Rose},~\bfnm{Sherri}\binits{S.}},
  \bauthor{\bsnm{Joshi},~\bfnm{Shalmali}\binits{S.}},
  \bauthor{\bsnm{Ferryman},~\bfnm{Kadija}\binits{K.}} \AND
  \bauthor{\bsnm{Ghassemi},~\bfnm{Marzyeh}\binits{M.}}
(\byear{2020}).
\btitle{Ethical Machine Learning in Health}.
\bjournal{arXiv preprint arXiv:2009.10576}.
\end{barticle}
\endbibitem

\bibitem[\protect\citeauthoryear{Chipman, George and
  McCulloch}{2007}]{chipman2007}
\begin{binproceedings}[author]
\bauthor{\bsnm{Chipman},~\bfnm{Hugh~A.}\binits{H.~A.}},
  \bauthor{\bsnm{George},~\bfnm{Edward~I.}\binits{E.~I.}} \AND
  \bauthor{\bsnm{McCulloch},~\bfnm{Robert}\binits{R.}}
(\byear{2007}).
\btitle{{Bayesian ensemble learning}}.
In \bbooktitle{{Advances in Neural Information Processing Systems 19 -
  Proceedings of the 2006 Conference}}
\bpages{265--272}.
\end{binproceedings}
\endbibitem

\bibitem[\protect\citeauthoryear{Chipman, George and
  McCulloch}{2010}]{chipman2010}
\begin{barticle}[author]
\bauthor{\bsnm{Chipman},~\bfnm{Hugh~A.}\binits{H.~A.}},
  \bauthor{\bsnm{George},~\bfnm{Edward~I.}\binits{E.~I.}} \AND
  \bauthor{\bsnm{McCulloch},~\bfnm{Robert~E.}\binits{R.~E.}}
(\byear{2010}).
\btitle{{{BART}}: Bayesian Additive Regression Trees}.
\bjournal{The Annals of Applied Statistics}
\bvolume{4}
\bpages{266--298}.
\bmrnumber{MR2758172}
\end{barticle}
\endbibitem

\bibitem[\protect\citeauthoryear{Cole and Stuart}{2010}]{cole2010}
\begin{barticle}[author]
\bauthor{\bsnm{Cole},~\bfnm{S.~R.}\binits{S.~R.}} \AND
  \bauthor{\bsnm{Stuart},~\bfnm{E.~A.}\binits{E.~A.}}
(\byear{2010}).
\btitle{Generalizing Evidence from Randomized Clinical Trials to Target
  Populations: The {{ACTG}} 320 Trial}.
\bjournal{American Journal of Epidemiology}
\bvolume{172}
\bpages{107--115}.
\bdoi{10.1093/aje/kwq084}
\end{barticle}
\endbibitem

\bibitem[\protect\citeauthoryear{Colnet et~al.}{2020}]{colnet2020}
\begin{barticle}[author]
\bauthor{\bsnm{Colnet},~\bfnm{B{\'e}n{\'e}dicte}\binits{B.}},
  \bauthor{\bsnm{Mayer},~\bfnm{Imke}\binits{I.}},
  \bauthor{\bsnm{Chen},~\bfnm{Guanhua}\binits{G.}},
  \bauthor{\bsnm{Dieng},~\bfnm{Awa}\binits{A.}},
  \bauthor{\bsnm{Li},~\bfnm{Ruohong}\binits{R.}},
  \bauthor{\bsnm{Varoquaux},~\bfnm{Ga{\"e}l}\binits{G.}},
  \bauthor{\bsnm{Vert},~\bfnm{Jean-Philippe}\binits{J.-P.}},
  \bauthor{\bsnm{Josse},~\bfnm{Julie}\binits{J.}} \AND
  \bauthor{\bsnm{Yang},~\bfnm{Shu}\binits{S.}}
(\byear{2020}).
\btitle{Causal Inference Methods for Combining Randomized Trials and
  Observational Studies: A Review}.
\bjournal{arXiv:2011.08047 [stat]}.
\end{barticle}
\endbibitem

\bibitem[\protect\citeauthoryear{Correa and Bareinboim}{2017}]{correa2017}
\begin{binproceedings}[author]
\bauthor{\bsnm{Correa},~\bfnm{Juan~D.}\binits{J.~D.}} \AND
  \bauthor{\bsnm{Bareinboim},~\bfnm{Elias}\binits{E.}}
(\byear{2017}).
\btitle{Causal Effect Identification by Adjustment under Confounding and
  Selection Biases}.
In \bbooktitle{Proceedings of the {{Thirty}}-{{First AAAI Conference}} on
  {{Artificial Intelligence}}}.
\bseries{{{AAAI}}'17}
\bpages{3740--3746}.
\bpublisher{{AAAI Press}}.
\end{binproceedings}
\endbibitem

\bibitem[\protect\citeauthoryear{Correa, Tian and
  Bareinboim}{2018}]{correa2018}
\begin{binproceedings}[author]
\bauthor{\bsnm{Correa},~\bfnm{Juan~D.}\binits{J.~D.}},
  \bauthor{\bsnm{Tian},~\bfnm{Jin}\binits{J.}} \AND
  \bauthor{\bsnm{Bareinboim},~\bfnm{Elias}\binits{E.}}
(\byear{2018}).
\btitle{Generalized Adjustment under Confounding and Selection Biases}.
In \bbooktitle{{{AAAI}}}.
\end{binproceedings}
\endbibitem

\bibitem[\protect\citeauthoryear{Cronbach and Shapiro}{1982}]{cronbach1982}
\begin{bbook}[author]
\bauthor{\bsnm{Cronbach},~\bfnm{Lee~J.}\binits{L.~J.}} \AND
  \bauthor{\bsnm{Shapiro},~\bfnm{Karen}\binits{K.}}
(\byear{1982}).
\btitle{Designing Evaluations of Educational and Social Programs},
\bedition{1st} ed.
\bseries{A {{Joint Publication}} in the {{Jossey}}-{{Bass Series}} in
  {{Social}} and {{Behavioral Science}} \& in {{Higher Education}}}.
\bpublisher{{Jossey-Bass}}, \baddress{{San Francisco}}.
\end{bbook}
\endbibitem

\bibitem[\protect\citeauthoryear{Crump et~al.}{2008}]{crump2008}
\begin{barticle}[author]
\bauthor{\bsnm{Crump},~\bfnm{Richard~K.}\binits{R.~K.}},
  \bauthor{\bsnm{Hotz},~\bfnm{V.~Joseph}\binits{V.~J.}},
  \bauthor{\bsnm{Imbens},~\bfnm{Guido~W.}\binits{G.~W.}} \AND
  \bauthor{\bsnm{Mitnik},~\bfnm{Oscar~A.}\binits{O.~A.}}
(\byear{2008}).
\btitle{Nonparametric Tests for Treatment Effect Heterogeneity}.
\bjournal{Review of Economics and Statistics}
\bvolume{90}
\bpages{389--405}.
\bdoi{10.1162/rest.90.3.389}
\end{barticle}
\endbibitem

\bibitem[\protect\citeauthoryear{Dahabreh et~al.}{2017}]{dahabreh2017}
\begin{barticle}[author]
\bauthor{\bsnm{Dahabreh},~\bfnm{Issa}\binits{I.}},
  \bauthor{\bsnm{Robertson},~\bfnm{Sarah}\binits{S.}},
  \bauthor{\bsnm{Stuart},~\bfnm{Elizabeth}\binits{E.}} \AND
  \bauthor{\bsnm{Hernan},~\bfnm{Miguel}\binits{M.}}
(\byear{2017}).
\btitle{Extending Inferences from Randomized Participants to All Eligible
  Individuals Using Trials Nested within Cohort Studies}.
\bjournal{arXiv:1709.04589 [stat]}.
\end{barticle}
\endbibitem

\bibitem[\protect\citeauthoryear{Dahabreh et~al.}{2018}]{dahabreh2018}
\begin{barticle}[author]
\bauthor{\bsnm{Dahabreh},~\bfnm{Issa~J.}\binits{I.~J.}},
  \bauthor{\bsnm{Robertson},~\bfnm{Sarah~E.}\binits{S.~E.}},
  \bauthor{\bsnm{Steingrimsson},~\bfnm{Jon~A.}\binits{J.~A.}},
  \bauthor{\bsnm{Stuart},~\bfnm{Elizabeth~A.}\binits{E.~A.}} \AND
  \bauthor{\bsnm{Hernan},~\bfnm{Miguel~A.}\binits{M.~A.}}
(\byear{2018}).
\btitle{Extending Inferences from a Randomized Trial to a New Target
  Population}.
\bjournal{arXiv:1805.00550 [stat]}.
\end{barticle}
\endbibitem

\bibitem[\protect\citeauthoryear{Dahabreh et~al.}{2019a}]{dahabreh2019a}
\begin{barticle}[author]
\bauthor{\bsnm{Dahabreh},~\bfnm{Issa~J.}\binits{I.~J.}},
  \bauthor{\bsnm{Robertson},~\bfnm{Sarah~E.}\binits{S.~E.}},
  \bauthor{\bsnm{Tchetgen},~\bfnm{Eric~J.}\binits{E.~J.}},
  \bauthor{\bsnm{Stuart},~\bfnm{Elizabeth~A.}\binits{E.~A.}} \AND
  \bauthor{\bsnm{Hern{\'a}n},~\bfnm{Miguel~A.}\binits{M.~A.}}
(\byear{2019}a).
\btitle{Generalizing Causal Inferences from Individuals in Randomized Trials to
  All Trial-Eligible Individuals}.
\bjournal{Biometrics}
\bvolume{75}
\bpages{685--694}.
\bdoi{10.1111/biom.13009}
\end{barticle}
\endbibitem

\bibitem[\protect\citeauthoryear{Dahabreh et~al.}{2019b}]{dahabreh2019c}
\begin{barticle}[author]
\bauthor{\bsnm{Dahabreh},~\bfnm{Issa~J}\binits{I.~J.}},
  \bauthor{\bsnm{Robins},~\bfnm{James~M}\binits{J.~M.}},
  \bauthor{\bsnm{Haneuse},~\bfnm{Sebastien J-P.~A}\binits{S.~J.-P.~A.}},
  \bauthor{\bsnm{Saeed},~\bfnm{Iman}\binits{I.}},
  \bauthor{\bsnm{Robertson},~\bfnm{Sarah~E}\binits{S.~E.}},
  \bauthor{\bsnm{Stuart},~\bfnm{Elisabeth~A}\binits{E.~A.}} \AND
  \bauthor{\bsnm{Hern{\'a}n},~\bfnm{Miguel~A}\binits{M.~A.}}
(\byear{2019}b).
\btitle{Sensitivity Analysis Using Bias Functions for Studies Extending
  Inferences from a Randomized Trial to a Target Population}.
\end{barticle}
\endbibitem

\bibitem[\protect\citeauthoryear{Dahabreh et~al.}{2019c}]{dahabreh2019b}
\begin{barticle}[author]
\bauthor{\bsnm{Dahabreh},~\bfnm{Issa~J}\binits{I.~J.}},
  \bauthor{\bsnm{Hernan},~\bfnm{Miguel~A}\binits{M.~A.}},
  \bauthor{\bsnm{Robertson},~\bfnm{Sarah~E}\binits{S.~E.}},
  \bauthor{\bsnm{Buchanan},~\bfnm{Ashley}\binits{A.}} \AND
  \bauthor{\bsnm{Steingrimsson},~\bfnm{Jon~A}\binits{J.~A.}}
(\byear{2019}c).
\btitle{Generalizing Trial Findings Using Nested Trial Designs with
  Sub-Sampling of Non-Randomized Individuals}.
\end{barticle}
\endbibitem

\bibitem[\protect\citeauthoryear{Dahabreh et~al.}{2019d}]{dahabreh2019}
\begin{barticle}[author]
\bauthor{\bsnm{Dahabreh},~\bfnm{Issa~J}\binits{I.~J.}},
  \bauthor{\bsnm{Robertson},~\bfnm{Sarah~E}\binits{S.~E.}},
  \bauthor{\bsnm{Petito},~\bfnm{Lucia~C}\binits{L.~C.}},
  \bauthor{\bsnm{Hern{\'a}n},~\bfnm{Miguel~A}\binits{M.~A.}} \AND
  \bauthor{\bsnm{Steingrimsson},~\bfnm{Jon~A}\binits{J.~A.}}
(\byear{2019}d).
\btitle{Efficient and Robust Methods for Causally Interpretable Meta-Analysis:
  Transporting Inferences from Multiple Randomized Trials to a Target
  Population}.
\end{barticle}
\endbibitem

\bibitem[\protect\citeauthoryear{Davis}{1988}]{davis1988}
\begin{barticle}[author]
\bauthor{\bsnm{Davis},~\bfnm{K}\binits{K.}}
(\byear{1988}).
\btitle{The Comprehensive Cohort Study: The Use of Registry Data to Confirm and
  Extend a Randomized Trial}.
\bjournal{Recent results in cancer research}
\bvolume{111}
\bpages{138}.
\end{barticle}
\endbibitem

\bibitem[\protect\citeauthoryear{Dekkers et~al.}{2010}]{dekkers2010}
\begin{barticle}[author]
\bauthor{\bsnm{Dekkers},~\bfnm{O~M}\binits{O.~M.}}, \bauthor{\bsnm{{von
  Elm}},~\bfnm{E}\binits{E.}}, \bauthor{\bsnm{Algra},~\bfnm{A}\binits{A.}},
  \bauthor{\bsnm{Romijn},~\bfnm{J~A}\binits{J.~A.}} \AND
  \bauthor{\bsnm{Vandenbroucke},~\bfnm{J~P}\binits{J.~P.}}
(\byear{2010}).
\btitle{How to Assess the External Validity of Therapeutic Trials: A Conceptual
  Approach}.
\bjournal{International Journal of Epidemiology}
\bvolume{39}
\bpages{89--94}.
\bdoi{10.1093/ije/dyp174}
\end{barticle}
\endbibitem

\bibitem[\protect\citeauthoryear{Ding, Feller and Miratrix}{2016}]{ding2016}
\begin{barticle}[author]
\bauthor{\bsnm{Ding},~\bfnm{Peng}\binits{P.}},
  \bauthor{\bsnm{Feller},~\bfnm{Avi}\binits{A.}} \AND
  \bauthor{\bsnm{Miratrix},~\bfnm{Luke}\binits{L.}}
(\byear{2016}).
\btitle{Randomization Inference for Treatment Effect Variation}.
\bjournal{Journal of the Royal Statistical Society. Series B, Statistical
  methodology}
\bvolume{78}
\bpages{655--671}.
\end{barticle}
\endbibitem

\bibitem[\protect\citeauthoryear{Dong et~al.}{2020}]{dong2020}
\begin{barticle}[author]
\bauthor{\bsnm{Dong},~\bfnm{Nianbo}\binits{N.}},
  \bauthor{\bsnm{Stuart},~\bfnm{Elizabeth~A}\binits{E.~A.}},
  \bauthor{\bsnm{Lenis},~\bfnm{David}\binits{D.}} \AND
  \bauthor{\bsnm{Quynh~Nguyen},~\bfnm{Trang}\binits{T.}}
(\byear{2020}).
\btitle{Using Propensity Score Analysis of Survey Data to Estimate Population
  Average Treatment Effects: A Case Study Comparing Different Methods}.
\bjournal{Evaluation review}
\bvolume{44}
\bpages{84--108}.
\end{barticle}
\endbibitem

\bibitem[\protect\citeauthoryear{Eddy}{1989}]{eddy1989}
\begin{barticle}[author]
\bauthor{\bsnm{Eddy},~\bfnm{David}\binits{D.}}
(\byear{1989}).
\btitle{The Confidence Profile Method: A Bayesian Method for Assessing Health
  Technologies}.
\bjournal{Operations Research}
\bvolume{37}
\bpages{210--228}.
\end{barticle}
\endbibitem

\bibitem[\protect\citeauthoryear{Elliott and Valliant}{2017}]{elliott2017}
\begin{barticle}[author]
\bauthor{\bsnm{Elliott},~\bfnm{Michael~R.}\binits{M.~R.}} \AND
  \bauthor{\bsnm{Valliant},~\bfnm{Richard}\binits{R.}}
(\byear{2017}).
\btitle{Inference for Nonprobability Samples}.
\bjournal{Statistical science}
\bvolume{32}
\bpages{249--264}.
\end{barticle}
\endbibitem

\bibitem[\protect\citeauthoryear{Fang}{2017}]{fang2017}
\begin{bbook}[author]
\bauthor{\bsnm{Fang},~\bfnm{Albert}\binits{A.}}
(\byear{2017}).
\btitle{10 Things to Know about Heterogeneous Treatment Effects}.
\end{bbook}
\endbibitem

\bibitem[\protect\citeauthoryear{Flores and Mitnik}{2013}]{flores2013}
\begin{barticle}[author]
\bauthor{\bsnm{Flores},~\bfnm{Carlos~A.}\binits{C.~A.}} \AND
  \bauthor{\bsnm{Mitnik},~\bfnm{Oscar~A.}\binits{O.~A.}}
(\byear{2013}).
\btitle{Comparing Treatments across Labor Markets: An Assessment of
  Nonexperimental Multiple-Treatment Strategies}.
\bjournal{The Review of Economics and Statistics}
\bvolume{95}
\bpages{1691--1707}.
\end{barticle}
\endbibitem

\bibitem[\protect\citeauthoryear{Ford and Norrie}{2016}]{ford2016}
\begin{barticle}[author]
\bauthor{\bsnm{Ford},~\bfnm{Ian}\binits{I.}} \AND
  \bauthor{\bsnm{Norrie},~\bfnm{John}\binits{J.}}
(\byear{2016}).
\btitle{Pragmatic Trials}.
\bjournal{New England Journal of Medicine}
\bvolume{375}
\bpages{454--463}.
\bdoi{10.1056/NEJMra1510059}
\end{barticle}
\endbibitem

\bibitem[\protect\citeauthoryear{Frangakis}{2009}]{frangakis2009}
\begin{barticle}[author]
\bauthor{\bsnm{Frangakis},~\bfnm{Constantine}\binits{C.}}
(\byear{2009}).
\btitle{The Calibration of Treatment Effects from Clinical Trials to Target
  Populations}.
\bjournal{Clinical Trials: Journal of the Society for Clinical Trials}
\bvolume{6}
\bpages{136--140}.
\bdoi{10.1177/1740774509103868}
\end{barticle}
\endbibitem

\bibitem[\protect\citeauthoryear{Gabler et~al.}{2009}]{gabler2009}
\begin{barticle}[author]
\bauthor{\bsnm{Gabler},~\bfnm{Nicole~B}\binits{N.~B.}},
  \bauthor{\bsnm{Duan},~\bfnm{Naihua}\binits{N.}},
  \bauthor{\bsnm{Liao},~\bfnm{Diana}\binits{D.}},
  \bauthor{\bsnm{Elmore},~\bfnm{Joann~G}\binits{J.~G.}},
  \bauthor{\bsnm{Ganiats},~\bfnm{Theodore~G}\binits{T.~G.}} \AND
  \bauthor{\bsnm{Kravitz},~\bfnm{Richard~L}\binits{R.~L.}}
(\byear{2009}).
\btitle{Dealing with Heterogeneity of Treatment Effects: Is the Literature up
  to the Challenge?}
\bjournal{Trials}
\bvolume{10}
\bpages{43--43}.
\end{barticle}
\endbibitem

\bibitem[\protect\citeauthoryear{Gail and Simon}{1985}]{gail1985}
\begin{barticle}[author]
\bauthor{\bsnm{Gail},~\bfnm{M}\binits{M.}} \AND
  \bauthor{\bsnm{Simon},~\bfnm{R}\binits{R.}}
(\byear{1985}).
\btitle{Testing for Qualitative Interactions between Treatment Effects and
  Patient Subsets}.
\bjournal{Biometrics}
\bvolume{41}
\bpages{361}.
\end{barticle}
\endbibitem

\bibitem[\protect\citeauthoryear{Gechter}{2015}]{gechter2015}
\begin{barticle}[author]
\bauthor{\bsnm{Gechter},~\bfnm{Michael}\binits{M.}}
(\byear{2015}).
\btitle{Generalizing the Results from Social Experiments: Theory and Evidence
  from Mexico and India}.
\bjournal{Department of Economics, Pennsylvania State University}
\bvolume{Unpublished manuscript}
\bpages{50}.
\end{barticle}
\endbibitem

\bibitem[\protect\citeauthoryear{Gelman and Little}{1997}]{gelman1997}
\begin{barticle}[author]
\bauthor{\bsnm{Gelman},~\bfnm{Andrew}\binits{A.}} \AND
  \bauthor{\bsnm{Little},~\bfnm{Thomas~C}\binits{T.~C.}}
(\byear{1997}).
\btitle{Poststratification into Many Categories Using Hierarchical Logistic
  Regression}.
\bjournal{Survey Methodology}
\bvolume{23}
\bpages{127--135}.
\end{barticle}
\endbibitem

\bibitem[\protect\citeauthoryear{Glauner et~al.}{2017}]{glauner2017}
\begin{barticle}[author]
\bauthor{\bsnm{Glauner},~\bfnm{Patrick}\binits{P.}},
  \bauthor{\bsnm{Migliosi},~\bfnm{Angelo}\binits{A.}},
  \bauthor{\bsnm{Meira},~\bfnm{Jorge}\binits{J.}},
  \bauthor{\bsnm{Valtchev},~\bfnm{Petko}\binits{P.}},
  \bauthor{\bsnm{State},~\bfnm{Radu}\binits{R.}} \AND
  \bauthor{\bsnm{Bettinger},~\bfnm{Franck}\binits{F.}}
(\byear{2017}).
\btitle{Is Big Data Sufficient for a Reliable Detection of Non-Technical
  Losses?}
\bjournal{arXiv:1702.03767 [cs]}.
\end{barticle}
\endbibitem

\bibitem[\protect\citeauthoryear{Green and Glasgow}{2006}]{green2006}
\begin{barticle}[author]
\bauthor{\bsnm{Green},~\bfnm{Lawrence~W}\binits{L.~W.}} \AND
  \bauthor{\bsnm{Glasgow},~\bfnm{Russell~E}\binits{R.~E.}}
(\byear{2006}).
\btitle{Evaluating the Relevance, Generalization, and Applicability of
  Research: Issues in External Validation and Translation Methodology}.
\bjournal{Evaluation \& the health professions}
\bvolume{29}
\bpages{126--153}.
\end{barticle}
\endbibitem

\bibitem[\protect\citeauthoryear{Green and Kern}{2012}]{green2012}
\begin{barticle}[author]
\bauthor{\bsnm{Green},~\bfnm{D.~P.}\binits{D.~P.}} \AND
  \bauthor{\bsnm{Kern},~\bfnm{H.~L.}\binits{H.~L.}}
(\byear{2012}).
\btitle{Modeling Heterogeneous Treatment Effects in Survey Experiments with
  Bayesian Additive Regression Trees}.
\bjournal{Public Opinion Quarterly}
\bvolume{76}
\bpages{491--511}.
\bdoi{10.1093/poq/nfs036}
\end{barticle}
\endbibitem

\bibitem[\protect\citeauthoryear{Greenhouse et~al.}{2008}]{greenhouse2008}
\begin{barticle}[author]
\bauthor{\bsnm{Greenhouse}},
  \bauthor{\bsnm{Kelleher},~\bfnm{Kelly}\binits{K.}},
  \bauthor{\bsnm{Seltman},~\bfnm{Howard}\binits{H.}} \AND
  \bauthor{\bsnm{Gardner},~\bfnm{William}\binits{W.}}
(\byear{2008}).
\btitle{Generalizing from Clinical Trial Data: A Case Study. the Risk of
  Suicidality among Pediatric Antidepressant Users.}
\bjournal{Statistics in Medicine}
\bvolume{27}
\bpages{1801--13}.
\bdoi{10.1002/sim.3218}
\end{barticle}
\endbibitem

\bibitem[\protect\citeauthoryear{Greenhouse et~al.}{2017}]{greenhouse2017}
\begin{bincollection}[author]
\bauthor{\bsnm{Greenhouse},~\bfnm{Joel~B}\binits{J.~B.}},
  \bauthor{\bsnm{Kaizar},~\bfnm{Eloise~E}\binits{E.~E.}},
  \bauthor{\bsnm{Anderson},~\bfnm{Heather~D.}\binits{H.~D.}},
  \bauthor{\bsnm{Bridge},~\bfnm{Jeffrey~A.}\binits{J.~A.}},
  \bauthor{\bsnm{Libby},~\bfnm{Anne~M.}\binits{A.~M.}},
  \bauthor{\bsnm{Valuck},~\bfnm{Robert}\binits{R.}} \AND
  \bauthor{\bsnm{Kelleher},~\bfnm{Kelly~J.}\binits{K.~J.}}
(\byear{2017}).
\btitle{Combining Information from Multiple Data Sources: An Introduction to
  Cross-Design Synthesis with a Case Study}.
In \bbooktitle{Methods in {{Comparative Effectiveness Research}}}
\bpages{223--246}.
\bpublisher{{Chapman and Hall/CRC}}.
\end{bincollection}
\endbibitem

\bibitem[\protect\citeauthoryear{Greenland}{2005}]{greenland2005}
\begin{barticle}[author]
\bauthor{\bsnm{Greenland},~\bfnm{S}\binits{S.}}
(\byear{2005}).
\btitle{Multiple-Bias Modelling for Analysis of Observational Data}.
\bjournal{Journal Of The Royal Statistical Society Series A}
\bvolume{168}
\bpages{267--291}.
\end{barticle}
\endbibitem

\bibitem[\protect\citeauthoryear{Gunter, Zhu and Murphy}{2011}]{gunter2011}
\begin{barticle}[author]
\bauthor{\bsnm{Gunter},~\bfnm{L.}\binits{L.}},
  \bauthor{\bsnm{Zhu},~\bfnm{J.}\binits{J.}} \AND
  \bauthor{\bsnm{Murphy},~\bfnm{S.~A.}\binits{S.~A.}}
(\byear{2011}).
\btitle{Variable Selection for Qualitative Interactions}.
\bjournal{Statistical Methodology}
\bvolume{8}
\bpages{42--55}.
\bdoi{10.1016/j.stamet.2009.05.003}
\end{barticle}
\endbibitem

\bibitem[\protect\citeauthoryear{Haneuse}{2016}]{haneuse2016}
\begin{barticle}[author]
\bauthor{\bsnm{Haneuse},~\bfnm{Sebastien}\binits{S.}}
(\byear{2016}).
\btitle{Distinguishing Selection Bias and Confounding Bias in Comparative
  Effectiveness Research}.
\bjournal{Medical care}
\bvolume{54}
\bpages{e23--e29}.
\end{barticle}
\endbibitem

\bibitem[\protect\citeauthoryear{Haneuse et~al.}{2009}]{haneuse2009}
\begin{barticle}[author]
\bauthor{\bsnm{Haneuse},~\bfnm{S.}\binits{S.}},
  \bauthor{\bsnm{Schildcrout},~\bfnm{J.}\binits{J.}},
  \bauthor{\bsnm{Crane},~\bfnm{P.}\binits{P.}},
  \bauthor{\bsnm{Sonnen},~\bfnm{J.}\binits{J.}},
  \bauthor{\bsnm{Breitner},~\bfnm{J.}\binits{J.}} \AND
  \bauthor{\bsnm{Larson},~\bfnm{E.}\binits{E.}}
(\byear{2009}).
\btitle{Adjustment for Selection Bias in Observational Studies with Application
  to the Analysis of Autopsy Data}.
\bjournal{Neuroepidemiology}
\bvolume{32}
\bpages{229--239}.
\bdoi{10.1159/000197389}
\end{barticle}
\endbibitem

\bibitem[\protect\citeauthoryear{Hartman et~al.}{2015}]{hartman2015}
\begin{barticle}[author]
\bauthor{\bsnm{Hartman},~\bfnm{Erin}\binits{E.}},
  \bauthor{\bsnm{Grieve},~\bfnm{Richard}\binits{R.}},
  \bauthor{\bsnm{Ramsahai},~\bfnm{Roland}\binits{R.}} \AND
  \bauthor{\bsnm{Sekhon},~\bfnm{Jasjeet~S.}\binits{J.~S.}}
(\byear{2015}).
\btitle{From Sample Average Treatment Effect to Population Average Treatment
  Effect on the Treated: Combining Experimental with Observational Studies to
  Estimate Population Treatment Effects}.
\bjournal{Journal of the Royal Statistical Society: Series A (Statistics in
  Society)}
\bvolume{178}
\bpages{757--778}.
\bdoi{10.1111/rssa.12094}
\end{barticle}
\endbibitem

\bibitem[\protect\citeauthoryear{He et~al.}{2016}]{he2016}
\begin{barticle}[author]
\bauthor{\bsnm{He},~\bfnm{Zhe}\binits{Z.}},
  \bauthor{\bsnm{Ryan},~\bfnm{Patrick}\binits{P.}},
  \bauthor{\bsnm{Hoxha},~\bfnm{Julia}\binits{J.}},
  \bauthor{\bsnm{Wang},~\bfnm{Shuang}\binits{S.}},
  \bauthor{\bsnm{Carini},~\bfnm{Simona}\binits{S.}},
  \bauthor{\bsnm{Sim},~\bfnm{Ida}\binits{I.}} \AND
  \bauthor{\bsnm{Weng},~\bfnm{Chunhua}\binits{C.}}
(\byear{2016}).
\btitle{Multivariate Analysis of the Population Representativeness of Related
  Clinical Studies}.
\bjournal{Journal of biomedical informatics}
\bvolume{60}
\bpages{66--76}.
\end{barticle}
\endbibitem

\bibitem[\protect\citeauthoryear{Heckman}{1979}]{heckman1979}
\begin{barticle}[author]
\bauthor{\bsnm{Heckman},~\bfnm{James~J.}\binits{J.~J.}}
(\byear{1979}).
\btitle{Sample Selection Bias as a Specification Error}.
\bjournal{Econometrica}
\bvolume{47}
\bpages{153--161}.
\end{barticle}
\endbibitem

\bibitem[\protect\citeauthoryear{Henderson, Varadhan and
  Weiss}{2017}]{henderson2017}
\begin{barticle}[author]
\bauthor{\bsnm{Henderson},~\bfnm{Nicholas~C.}\binits{N.~C.}},
  \bauthor{\bsnm{Varadhan},~\bfnm{Ravi}\binits{R.}} \AND
  \bauthor{\bsnm{Weiss},~\bfnm{Carlos~O.}\binits{C.~O.}}
(\byear{2017}).
\btitle{Cross-Design Synthesis for Extending the Applicability of Trial
  Evidence When Treatment Effect Is Heterogenous: Part {{II}}. Application and
  External Validation}.
\bjournal{Communications in Statistics: Case Studies, Data Analysis and
  Applications}
\bvolume{3}
\bpages{7--20}.
\bdoi{10.1080/23737484.2017.1398056}
\end{barticle}
\endbibitem

\bibitem[\protect\citeauthoryear{Hern{\'a}n et~al.}{2008}]{hernan2008}
\begin{barticle}[author]
\bauthor{\bsnm{Hern{\'a}n},~\bfnm{Miguel~A.}\binits{M.~A.}},
  \bauthor{\bsnm{Alonso},~\bfnm{Alvaro}\binits{A.}},
  \bauthor{\bsnm{Logan},~\bfnm{Roger}\binits{R.}},
  \bauthor{\bsnm{Grodstein},~\bfnm{Francine}\binits{F.}},
  \bauthor{\bsnm{Michels},~\bfnm{Karin~B.}\binits{K.~B.}},
  \bauthor{\bsnm{Willett},~\bfnm{Walter~C.}\binits{W.~C.}},
  \bauthor{\bsnm{Manson},~\bfnm{JoAnn~E.}\binits{J.~E.}} \AND
  \bauthor{\bsnm{Robins},~\bfnm{James~M.}\binits{J.~M.}}
(\byear{2008}).
\btitle{Observational Studies Analyzed like Randomized Experiments: An
  Application to Postmenopausal Hormone Therapy and Coronary Heart Disease}.
\bjournal{Epidemiology}
\bvolume{19}
\bpages{766--779}.
\bdoi{10.1097/EDE.0b013e3181875e61}
\end{barticle}
\endbibitem

\bibitem[\protect\citeauthoryear{Hill}{2011}]{hill2011}
\begin{barticle}[author]
\bauthor{\bsnm{Hill},~\bfnm{Jennifer~L.}\binits{J.~L.}}
(\byear{2011}).
\btitle{Bayesian Nonparametric Modeling for Causal Inference}.
\bjournal{Journal of Computational and Graphical Statistics}
\bvolume{20}
\bpages{217--240}.
\bdoi{10.1198/jcgs.2010.08162}
\end{barticle}
\endbibitem

\bibitem[\protect\citeauthoryear{Horvitz and Thompson}{1952}]{horvitz1952}
\begin{barticle}[author]
\bauthor{\bsnm{Horvitz},~\bfnm{D.~G.}\binits{D.~G.}} \AND
  \bauthor{\bsnm{Thompson},~\bfnm{D.~J.}\binits{D.~J.}}
(\byear{1952}).
\btitle{A Generalization of Sampling without Replacement from a Finite
  Universe}.
\bjournal{Journal of the American Statistical Association}
\bvolume{47}
\bpages{663--685}.
\bdoi{10.1080/01621459.1952.10483446}
\end{barticle}
\endbibitem

\bibitem[\protect\citeauthoryear{Hotz, Imbens and Mortimer}{2005}]{hotz2005}
\begin{barticle}[author]
\bauthor{\bsnm{Hotz},~\bfnm{V.~Joseph}\binits{V.~J.}},
  \bauthor{\bsnm{Imbens},~\bfnm{Guido~W}\binits{G.~W.}} \AND
  \bauthor{\bsnm{Mortimer},~\bfnm{Julie~H}\binits{J.~H.}}
(\byear{2005}).
\btitle{Predicting the Efficacy of Future Training Programs Using Past
  Experiences at Other Locations}.
\bjournal{Journal of econometrics}
\bvolume{125}
\bpages{241--270}.
\end{barticle}
\endbibitem

\bibitem[\protect\citeauthoryear{Imai, King and Stuart}{2008}]{imai2008}
\begin{barticle}[author]
\bauthor{\bsnm{Imai},~\bfnm{Kosuke}\binits{K.}},
  \bauthor{\bsnm{King},~\bfnm{Gary}\binits{G.}} \AND
  \bauthor{\bsnm{Stuart},~\bfnm{Elizabeth~A.}\binits{E.~A.}}
(\byear{2008}).
\btitle{Misunderstandings between Experimentalists and Observationalists about
  Causal Inference}.
\bjournal{Journal of the Royal Statistical Society: Series A (Statistics in
  Society)}
\bvolume{171}
\bpages{481--502}.
\end{barticle}
\endbibitem

\bibitem[\protect\citeauthoryear{Johansson et~al.}{2018}]{johansson2018}
\begin{barticle}[author]
\bauthor{\bsnm{Johansson},~\bfnm{Fredrik~D.}\binits{F.~D.}},
  \bauthor{\bsnm{Kallus},~\bfnm{Nathan}\binits{N.}},
  \bauthor{\bsnm{Shalit},~\bfnm{Uri}\binits{U.}} \AND
  \bauthor{\bsnm{Sontag},~\bfnm{David}\binits{D.}}
(\byear{2018}).
\btitle{Learning Weighted Representations for Generalization across Designs}.
\bjournal{arXiv:1802.08598 [stat]}.
\end{barticle}
\endbibitem

\bibitem[\protect\citeauthoryear{Josey et~al.}{2020a}]{josey2020}
\begin{barticle}[author]
\bauthor{\bsnm{Josey},~\bfnm{Kevin~P}\binits{K.~P.}},
  \bauthor{\bsnm{Yang},~\bfnm{Fan}\binits{F.}},
  \bauthor{\bsnm{Ghosh},~\bfnm{Debashis}\binits{D.}} \AND
  \bauthor{\bsnm{Raghavan},~\bfnm{Sridharan}\binits{S.}}
(\byear{2020}a).
\btitle{A Calibration Approach to Transportability with Observational Data}.
\end{barticle}
\endbibitem

\bibitem[\protect\citeauthoryear{Josey et~al.}{2020b}]{josey2020a}
\begin{barticle}[author]
\bauthor{\bsnm{Josey},~\bfnm{Kevin~P}\binits{K.~P.}},
  \bauthor{\bsnm{Berkowitz},~\bfnm{Seth~A}\binits{S.~A.}},
  \bauthor{\bsnm{Ghosh},~\bfnm{Debashis}\binits{D.}} \AND
  \bauthor{\bsnm{Raghavan},~\bfnm{Sridharan}\binits{S.}}
(\byear{2020}b).
\btitle{Transporting Experimental Results with Entropy Balancing}.
\end{barticle}
\endbibitem

\bibitem[\protect\citeauthoryear{Kaizar}{2011}]{kaizar2011}
\begin{barticle}[author]
\bauthor{\bsnm{Kaizar},~\bfnm{Eloise~E.}\binits{E.~E.}}
(\byear{2011}).
\btitle{Estimating Treatment Effect via Simple Cross Design Synthesis}.
\bjournal{Statistics in Medicine}
\bvolume{30}
\bpages{2986--3009}.
\bdoi{10.1002/sim.4339}
\end{barticle}
\endbibitem

\bibitem[\protect\citeauthoryear{Kaizar}{2015}]{kaizar2015}
\begin{barticle}[author]
\bauthor{\bsnm{Kaizar},~\bfnm{Eloise~E.}\binits{E.~E.}}
(\byear{2015}).
\btitle{Incorporating Both Randomized and Observational Data into a Single
  Analysis}.
\bjournal{Annual Review of Statistics and Its Application}
\bvolume{2}
\bpages{49--72}.
\bdoi{10.1146/annurev-statistics-010814-020249}
\end{barticle}
\endbibitem

\bibitem[\protect\citeauthoryear{Kallus, Puli and Shalit}{2018}]{kallus2018}
\begin{barticle}[author]
\bauthor{\bsnm{Kallus},~\bfnm{Nathan}\binits{N.}},
  \bauthor{\bsnm{Puli},~\bfnm{Aahlad~Manas}\binits{A.~M.}} \AND
  \bauthor{\bsnm{Shalit},~\bfnm{Uri}\binits{U.}}
(\byear{2018}).
\btitle{Removing Hidden Confounding by Experimental Grounding}.
\bjournal{arXiv:1810.11646 [cs, stat]}.
\end{barticle}
\endbibitem

\bibitem[\protect\citeauthoryear{Kennedy and Gelman}{2019}]{kennedy2019}
\begin{barticle}[author]
\bauthor{\bsnm{Kennedy},~\bfnm{Lauren}\binits{L.}} \AND
  \bauthor{\bsnm{Gelman},~\bfnm{Andrew}\binits{A.}}
(\byear{2019}).
\btitle{Know Your Population and Know Your Model: Using Model-Based Regression
  and Poststratification to Generalize Findings beyond the Observed Sample}.
\bjournal{arXiv:1906.11323 [stat]}.
\end{barticle}
\endbibitem

\bibitem[\protect\citeauthoryear{{Kennedy-Martin}
  et~al.}{2015}]{kennedy-martin2015}
\begin{barticle}[author]
\bauthor{\bsnm{{Kennedy-Martin}},~\bfnm{Tessa}\binits{T.}},
  \bauthor{\bsnm{Curtis},~\bfnm{Sarah}\binits{S.}},
  \bauthor{\bsnm{Faries},~\bfnm{Douglas}\binits{D.}},
  \bauthor{\bsnm{Robinson},~\bfnm{Susan}\binits{S.}} \AND
  \bauthor{\bsnm{Johnston},~\bfnm{Joseph}\binits{J.}}
(\byear{2015}).
\btitle{A Literature Review on the Representativeness of Randomized Controlled
  Trial Samples and Implications for the External Validity of Trial Results}.
\bjournal{Trials}
\bvolume{16}
\bpages{495--495}.
\end{barticle}
\endbibitem

\bibitem[\protect\citeauthoryear{Kern et~al.}{2016}]{kern2016}
\begin{barticle}[author]
\bauthor{\bsnm{Kern},~\bfnm{Holger~L.}\binits{H.~L.}},
  \bauthor{\bsnm{Stuart},~\bfnm{Elizabeth~A.}\binits{E.~A.}},
  \bauthor{\bsnm{Hill},~\bfnm{Jennifer}\binits{J.}} \AND
  \bauthor{\bsnm{Green},~\bfnm{Donald~P.}\binits{D.~P.}}
(\byear{2016}).
\btitle{Assessing Methods for Generalizing Experimental Impact Estimates to
  Target Populations}.
\bjournal{Journal of Research on Educational Effectiveness}
\bvolume{9}
\bpages{103--127}.
\bdoi{10.1080/19345747.2015.1060282}
\end{barticle}
\endbibitem

\bibitem[\protect\citeauthoryear{Kim et~al.}{2018}]{kim2018}
\begin{barticle}[author]
\bauthor{\bsnm{Kim},~\bfnm{Jae~Kwang}\binits{J.~K.}},
  \bauthor{\bsnm{Park},~\bfnm{Seho}\binits{S.}},
  \bauthor{\bsnm{Chen},~\bfnm{Yilin}\binits{Y.}} \AND
  \bauthor{\bsnm{Wu},~\bfnm{Changbao}\binits{C.}}
(\byear{2018}).
\btitle{Combining Non-Probability and Probability Survey Samples through Mass
  Imputation}.
\end{barticle}
\endbibitem

\bibitem[\protect\citeauthoryear{Lesko et~al.}{2017}]{lesko2017}
\begin{barticle}[author]
\bauthor{\bsnm{Lesko},~\bfnm{Catherine~R.}\binits{C.~R.}},
  \bauthor{\bsnm{Buchanan},~\bfnm{Ashley~L.}\binits{A.~L.}},
  \bauthor{\bsnm{Westreich},~\bfnm{Daniel}\binits{D.}},
  \bauthor{\bsnm{Edwards},~\bfnm{Jessie~K.}\binits{J.~K.}},
  \bauthor{\bsnm{Hudgens},~\bfnm{Michael~G.}\binits{M.~G.}} \AND
  \bauthor{\bsnm{Cole},~\bfnm{Stephen~R.}\binits{S.~R.}}
(\byear{2017}).
\btitle{Generalizing Study Results: A Potential Outcomes Perspective}.
\bjournal{Epidemiology}
\bvolume{28}
\bpages{553--561}.
\bdoi{10.1097/EDE.0000000000000664}
\end{barticle}
\endbibitem

\bibitem[\protect\citeauthoryear{Lipkovich et~al.}{2011}]{lipkovich2011}
\begin{barticle}[author]
\bauthor{\bsnm{Lipkovich},~\bfnm{Ilya}\binits{I.}},
  \bauthor{\bsnm{Dmitrienko},~\bfnm{Alex}\binits{A.}},
  \bauthor{\bsnm{Denne},~\bfnm{Jonathan}\binits{J.}} \AND
  \bauthor{\bsnm{Enas},~\bfnm{Gregory}\binits{G.}}
(\byear{2011}).
\btitle{Subgroup Identification Based on Differential Effect Search-a Recursive
  Partitioning Method for Establishing Response to Treatment in Patient
  Subpopulations: Subgroup Identification Based on Differential Effect Search
  ({{SIDES}})}.
\bjournal{Statistics in Medicine}
\bvolume{30}
\bpages{2601--21}.
\bdoi{10.1002/sim.4289}
\end{barticle}
\endbibitem

\bibitem[\protect\citeauthoryear{Lu et~al.}{2019}]{lu2019}
\begin{barticle}[author]
\bauthor{\bsnm{Lu},~\bfnm{Yi}\binits{Y.}},
  \bauthor{\bsnm{Scharfstein},~\bfnm{Daniel~O}\binits{D.~O.}},
  \bauthor{\bsnm{Brooks},~\bfnm{Maria~M}\binits{M.~M.}},
  \bauthor{\bsnm{Quach},~\bfnm{Kevin}\binits{K.}} \AND
  \bauthor{\bsnm{Kennedy},~\bfnm{Edward~H}\binits{E.~H.}}
(\byear{2019}).
\btitle{Causal Inference for Comprehensive Cohort Studies}.
\end{barticle}
\endbibitem

\bibitem[\protect\citeauthoryear{Luedtke, Carone and {van der
  Laan}}{2019}]{luedtke2019}
\begin{barticle}[author]
\bauthor{\bsnm{Luedtke},~\bfnm{Alex}\binits{A.}},
  \bauthor{\bsnm{Carone},~\bfnm{Marco}\binits{M.}} \AND \bauthor{\bsnm{{van der
  Laan}},~\bfnm{Mark~J}\binits{M.~J.}}
(\byear{2019}).
\btitle{An Omnibus Non-parametric Test of Equality in Distribution for Unknown
  Functions}.
\bjournal{Journal of the Royal Statistical Society. Series B, Statistical
  methodology}
\bvolume{81}
\bpages{75--99}.
\end{barticle}
\endbibitem

\bibitem[\protect\citeauthoryear{Lunceford and Davidian}{2004}]{lunceford2004}
\begin{barticle}[author]
\bauthor{\bsnm{Lunceford},~\bfnm{Jared~K}\binits{J.~K.}} \AND
  \bauthor{\bsnm{Davidian},~\bfnm{Marie}\binits{M.}}
(\byear{2004}).
\btitle{Stratification and Weighting via the Propensity Score in Estimation of
  Causal Treatment Effects: A Comparative Study.}
\bjournal{Statistics in Medicine}
\bvolume{23}
\bpages{2937--60}.
\bdoi{10.1002/sim.1903}
\end{barticle}
\endbibitem

\bibitem[\protect\citeauthoryear{Marcus}{1997}]{marcus1997}
\begin{barticle}[author]
\bauthor{\bsnm{Marcus},~\bfnm{SM}\binits{S.}}
(\byear{1997}).
\btitle{Assessing Non-Consent Bias with Parallel Randomized and Nonrandomized
  Clinical Trials}.
\bjournal{Journal Of Clinical Epidemiology}
\bvolume{50}
\bpages{823--828}.
\end{barticle}
\endbibitem

\bibitem[\protect\citeauthoryear{Miettinen}{1972}]{miettinen1972}
\begin{barticle}[author]
\bauthor{\bsnm{Miettinen},~\bfnm{Olli~S.}\binits{O.~S.}}
(\byear{1972}).
\btitle{Standardization of Risk Ratios}.
\bjournal{American Journal of Epidemiology}
\bvolume{96}
\bpages{383--388}.
\bdoi{10.1093/oxfordjournals.aje.a121470}
\end{barticle}
\endbibitem

\bibitem[\protect\citeauthoryear{{Moreno-Torres}
  et~al.}{2012}]{moreno-torres2012}
\begin{barticle}[author]
\bauthor{\bsnm{{Moreno-Torres}},~\bfnm{Jose~G.}\binits{J.~G.}},
  \bauthor{\bsnm{Raeder},~\bfnm{Troy}\binits{T.}},
  \bauthor{\bsnm{{Alaiz-Rodr{\'i}guez}},~\bfnm{Roc{\'i}o}\binits{R.}},
  \bauthor{\bsnm{Chawla},~\bfnm{Nitesh~V.}\binits{N.~V.}} \AND
  \bauthor{\bsnm{Herrera},~\bfnm{Francisco}\binits{F.}}
(\byear{2012}).
\btitle{A Unifying View on Dataset Shift in Classification}.
\bjournal{Pattern Recognition}
\bvolume{45}
\bpages{521--530}.
\bdoi{10.1016/j.patcog.2011.06.019}
\end{barticle}
\endbibitem

\bibitem[\protect\citeauthoryear{Neugebauer and {van der
  Laan}}{2005}]{neugebauer2005}
\begin{barticle}[author]
\bauthor{\bsnm{Neugebauer},~\bfnm{Romain}\binits{R.}} \AND \bauthor{\bsnm{{van
  der Laan}},~\bfnm{Mark}\binits{M.}}
(\byear{2005}).
\btitle{Why Prefer Double Robust Estimators in Causal Inference?}
\bjournal{Journal of statistical planning and inference}
\bvolume{129}
\bpages{405--426}.
\end{barticle}
\endbibitem

\bibitem[\protect\citeauthoryear{Nguyen et~al.}{2017}]{nguyen2017}
\begin{barticle}[author]
\bauthor{\bsnm{Nguyen},~\bfnm{Trang~Quynh}\binits{T.~Q.}},
  \bauthor{\bsnm{Ebnesajjad},~\bfnm{Cyrus}\binits{C.}},
  \bauthor{\bsnm{Cole},~\bfnm{Stephen~R.}\binits{S.~R.}} \AND
  \bauthor{\bsnm{Stuart},~\bfnm{Elizabeth~A.}\binits{E.~A.}}
(\byear{2017}).
\btitle{Sensitivity Analysis for an Unobserved Moderator in
  {{RCT}}-to-Target-Population Generalization of Treatment Effects}.
\bjournal{Annals of Applied Statistics}
\bvolume{11}
\bpages{225--247}.
\bdoi{10.1214/16-AOAS1001}
\bmrnumber{MR3634322}
\end{barticle}
\endbibitem

\bibitem[\protect\citeauthoryear{Nguyen et~al.}{2018}]{nguyen2018}
\begin{barticle}[author]
\bauthor{\bsnm{Nguyen},~\bfnm{Trang~Quynh}\binits{T.~Q.}},
  \bauthor{\bsnm{Ackerman},~\bfnm{Benjamin}\binits{B.}},
  \bauthor{\bsnm{Schmid},~\bfnm{Ian}\binits{I.}},
  \bauthor{\bsnm{Cole},~\bfnm{Stephen~R.}\binits{S.~R.}} \AND
  \bauthor{\bsnm{Stuart},~\bfnm{Elizabeth~A.}\binits{E.~A.}}
(\byear{2018}).
\btitle{Sensitivity Analyses for Effect Modifiers Not Observed in the Target
  Population When Generalizing Treatment Effects from a Randomized Controlled
  Trial: Assumptions, Models, Effect Scales, Data Scenarios, and Implementation
  Details}.
\bjournal{PLOS ONE}
\bvolume{13}
\bpages{e0208795}.
\bdoi{10.1371/journal.pone.0208795}
\end{barticle}
\endbibitem

\bibitem[\protect\citeauthoryear{Nie et~al.}{2013}]{nie2013}
\begin{barticle}[author]
\bauthor{\bsnm{Nie},~\bfnm{Lei}\binits{L.}},
  \bauthor{\bsnm{Zhang},~\bfnm{Zhiwei}\binits{Z.}},
  \bauthor{\bsnm{Rubin},~\bfnm{Daniel}\binits{D.}} \AND
  \bauthor{\bsnm{Chu},~\bfnm{Jianxiong}\binits{J.}}
(\byear{2013}).
\btitle{Likelihood Reweighting Methods to Reduce Potential Bias in
  Noninferiority Trials Which Rely on Historical Data to Make Inference}.
\bjournal{The Annals of Applied Statistics}
\bvolume{7}
\bpages{1796--1813}.
\bdoi{10.1214/13-AOAS655}
\end{barticle}
\endbibitem

\bibitem[\protect\citeauthoryear{O'~Muircheartaigh and
  Hedges}{2014}]{omuircheartaigh2014}
\begin{barticle}[author]
\bauthor{\bsnm{O'~Muircheartaigh},~\bfnm{Colm}\binits{C.}} \AND
  \bauthor{\bsnm{Hedges},~\bfnm{Larry~V.}\binits{L.~V.}}
(\byear{2014}).
\btitle{Generalizing from Unrepresentative Experiments: A Stratified Propensity
  Score Approach}.
\bjournal{Journal of the Royal Statistical Society: Series C (Applied
  Statistics)}
\bvolume{63}
\bpages{195--210}.
\end{barticle}
\endbibitem

\bibitem[\protect\citeauthoryear{Olsen et~al.}{2013}]{olsen2013}
\begin{barticle}[author]
\bauthor{\bsnm{Olsen},~\bfnm{Robert~B}\binits{R.~B.}},
  \bauthor{\bsnm{Orr},~\bfnm{Larry~L}\binits{L.~L.}},
  \bauthor{\bsnm{Bell},~\bfnm{Stephen~H}\binits{S.~H.}} \AND
  \bauthor{\bsnm{Stuart},~\bfnm{Elizabeth~A}\binits{E.~A.}}
(\byear{2013}).
\btitle{External Validity in Policy Evaluations That Choose Sites Purposively:
  External Validity in Policy Evaluations}.
\bjournal{Journal of policy analysis and management}
\bvolume{32}
\bpages{107--121}.
\end{barticle}
\endbibitem

\bibitem[\protect\citeauthoryear{Pan and Schaubel}{2008}]{pan2008}
\begin{barticle}[author]
\bauthor{\bsnm{Pan},~\bfnm{Qing}\binits{Q.}} \AND
  \bauthor{\bsnm{Schaubel},~\bfnm{Douglas~E.}\binits{D.~E.}}
(\byear{2008}).
\btitle{Proportional Hazards Models Based on Biased Samples and Estimated
  Selection Probabilities}.
\bjournal{The Canadian Journal of Statistics / La Revue Canadienne de
  Statistique}
\bvolume{36}
\bpages{111--127}.
\end{barticle}
\endbibitem

\bibitem[\protect\citeauthoryear{Pan and Schaubel}{2009}]{pan2009}
\begin{barticle}[author]
\bauthor{\bsnm{Pan},~\bfnm{Qing}\binits{Q.}} \AND
  \bauthor{\bsnm{Schaubel},~\bfnm{Douglas~E.}\binits{D.~E.}}
(\byear{2009}).
\btitle{Evaluating Bias Correction in Weighted Proportional Hazards
  Regression}.
\bjournal{Lifetime Data Analysis}
\bvolume{15}
\bpages{120--146}.
\bdoi{10.1007/s10985-008-9102-4}
\end{barticle}
\endbibitem

\bibitem[\protect\citeauthoryear{Park, Gelman and Bafumi}{2004}]{park2004}
\begin{barticle}[author]
\bauthor{\bsnm{Park},~\bfnm{David~K}\binits{D.~K.}},
  \bauthor{\bsnm{Gelman},~\bfnm{Andrew}\binits{A.}} \AND
  \bauthor{\bsnm{Bafumi},~\bfnm{Joseph}\binits{J.}}
(\byear{2004}).
\btitle{Bayesian Multilevel Estimation with Poststratification: State-Level
  Estimates from National Polls}.
\bjournal{Political Analysis}
\bvolume{12}
\bpages{375--385}.
\end{barticle}
\endbibitem

\bibitem[\protect\citeauthoryear{Pearl}{2000}]{pearl2000}
\begin{bbook}[author]
\bauthor{\bsnm{Pearl},~\bfnm{Judea}\binits{J.}}
(\byear{2000}).
\btitle{Causality : Models, Reasoning, and Inference}.
\bpublisher{{Cambridge University Press}}.
\end{bbook}
\endbibitem

\bibitem[\protect\citeauthoryear{Pearl}{2015}]{pearl2015}
\begin{barticle}[author]
\bauthor{\bsnm{Pearl},~\bfnm{Judea}\binits{J.}}
(\byear{2015}).
\btitle{Generalizing Experimental Findings}.
\bjournal{Journal of Causal Inference}
\bvolume{3}.
\bdoi{10.1515/jci-2015-0025}
\end{barticle}
\endbibitem

\bibitem[\protect\citeauthoryear{Pearl and Bareinboim}{2011}]{pearl2011}
\begin{binproceedings}[author]
\bauthor{\bsnm{Pearl},~\bfnm{Judea}\binits{J.}} \AND
  \bauthor{\bsnm{Bareinboim},~\bfnm{Elias}\binits{E.}}
(\byear{2011}).
\btitle{Transportability of Causal and Statistical Relations: A Formal
  Approach}.
In \bbooktitle{2011 {{IEEE}} 11th {{International Conference}} on {{Data Mining
  Workshops}}}
\bpages{540--547}.
\bpublisher{{IEEE}}, \baddress{{Vancouver, BC, Canada}}.
\bdoi{10.1109/ICDMW.2011.169}
\end{binproceedings}
\endbibitem

\bibitem[\protect\citeauthoryear{Pearl and Bareinboim}{2014}]{pearl2014}
\begin{barticle}[author]
\bauthor{\bsnm{Pearl},~\bfnm{Judea}\binits{J.}} \AND
  \bauthor{\bsnm{Bareinboim},~\bfnm{Elias}\binits{E.}}
(\byear{2014}).
\btitle{External Validity: From Do-Calculus to Transportability across
  Populations}.
\bjournal{Statistical Science}
\bvolume{29}
\bpages{579--595}.
\bdoi{10.1214/14-STS486}
\end{barticle}
\endbibitem

\bibitem[\protect\citeauthoryear{Phillippo et~al.}{2018}]{phillippo2018}
\begin{barticle}[author]
\bauthor{\bsnm{Phillippo},~\bfnm{David~M}\binits{D.~M.}},
  \bauthor{\bsnm{Ades},~\bfnm{Anthony~E}\binits{A.~E.}},
  \bauthor{\bsnm{Dias},~\bfnm{Sofia}\binits{S.}},
  \bauthor{\bsnm{Palmer},~\bfnm{Stephen}\binits{S.}},
  \bauthor{\bsnm{Abrams},~\bfnm{Keith~R}\binits{K.~R.}} \AND
  \bauthor{\bsnm{Welton},~\bfnm{Nicky~J}\binits{N.~J.}}
(\byear{2018}).
\btitle{Methods for Population-Adjusted Indirect Comparisons in Health
  Technology Appraisal}.
\bjournal{Medical decision making}
\bvolume{38}
\bpages{200--211}.
\end{barticle}
\endbibitem

\bibitem[\protect\citeauthoryear{Pool, Abelson and Popkin}{1964}]{pool1964}
\begin{bbook}[author]
\bauthor{\bsnm{Pool},~\bfnm{I}\binits{I.}},
  \bauthor{\bsnm{Abelson},~\bfnm{R}\binits{R.}} \AND
  \bauthor{\bsnm{Popkin},~\bfnm{S}\binits{S.}}
(\byear{1964}).
\btitle{Candidates, Issues and Strategies; a Computer Simulation of the 1960
  Presidential Election}.
\bpublisher{{Massachusetts Institute of Technology Press}}.
\end{bbook}
\endbibitem

\bibitem[\protect\citeauthoryear{Prentice et~al.}{2005}]{prentice2005}
\begin{barticle}[author]
\bauthor{\bsnm{Prentice},~\bfnm{Ross~L.}\binits{R.~L.}},
  \bauthor{\bsnm{Langer},~\bfnm{Robert}\binits{R.}},
  \bauthor{\bsnm{Stefanick},~\bfnm{Marcia~L.}\binits{M.~L.}},
  \bauthor{\bsnm{Howard},~\bfnm{Barbara~V.}\binits{B.~V.}},
  \bauthor{\bsnm{Pettinger},~\bfnm{Mary}\binits{M.}},
  \bauthor{\bsnm{Anderson},~\bfnm{Garnet}\binits{G.}},
  \bauthor{\bsnm{Barad},~\bfnm{David}\binits{D.}},
  \bauthor{\bsnm{Curb},~\bfnm{J.~David}\binits{J.~D.}},
  \bauthor{\bsnm{Kotchen},~\bfnm{Jane}\binits{J.}},
  \bauthor{\bsnm{Kuller},~\bfnm{Lewis}\binits{L.}},
  \bauthor{\bsnm{Limacher},~\bfnm{Marian}\binits{M.}} \AND
  \bauthor{\bsnm{{Wactawski-Wende}},~\bfnm{Jean}\binits{J.}}
(\byear{2005}).
\btitle{Combined Postmenopausal Hormone Therapy and Cardiovascular Disease:
  Toward Resolving the Discrepancy between Observational Studies and the
  Women's Health Initiative Clinical Trial}.
\bjournal{American Journal of Epidemiology}
\bvolume{162}
\bpages{404--414}.
\bdoi{10.1093/aje/kwi223}
\end{barticle}
\endbibitem

\bibitem[\protect\citeauthoryear{Prevost, Abrams and Jones}{2000}]{prevost2000}
\begin{barticle}[author]
\bauthor{\bsnm{Prevost},~\bfnm{Teresa~C.}\binits{T.~C.}},
  \bauthor{\bsnm{Abrams},~\bfnm{Keith~R.}\binits{K.~R.}} \AND
  \bauthor{\bsnm{Jones},~\bfnm{David~R.}\binits{D.~R.}}
(\byear{2000}).
\btitle{Hierarchical Models in Generalized Synthesis of Evidence: An Example
  Based on Studies of Breast Cancer Screening}.
\bjournal{Statistics in Medicine}
\bvolume{19}
\bpages{3359--3376}.
\end{barticle}
\endbibitem

\bibitem[\protect\citeauthoryear{Qian, Chakraborty and Maiti}{2019}]{qian2019}
\begin{barticle}[author]
\bauthor{\bsnm{Qian},~\bfnm{Min}\binits{M.}},
  \bauthor{\bsnm{Chakraborty},~\bfnm{Bibhas}\binits{B.}} \AND
  \bauthor{\bsnm{Maiti},~\bfnm{Raju}\binits{R.}}
(\byear{2019}).
\btitle{A Sequential Significance Test for Treatment by Covariate
  Interactions}.
\bjournal{arXiv:1901.08738 [stat]}.
\end{barticle}
\endbibitem

\bibitem[\protect\citeauthoryear{Rosenman et~al.}{2018}]{rosenman2018}
\begin{barticle}[author]
\bauthor{\bsnm{Rosenman},~\bfnm{Evan}\binits{E.}},
  \bauthor{\bsnm{Owen},~\bfnm{Art~B.}\binits{A.~B.}},
  \bauthor{\bsnm{Baiocchi},~\bfnm{Michael}\binits{M.}} \AND
  \bauthor{\bsnm{Banack},~\bfnm{Hailey}\binits{H.}}
(\byear{2018}).
\btitle{Propensity Score Methods for Merging Observational and Experimental
  Datasets}.
\bjournal{arXiv:1804.07863 [stat]}.
\end{barticle}
\endbibitem

\bibitem[\protect\citeauthoryear{Rosenman et~al.}{2020}]{rosenman2020}
\begin{barticle}[author]
\bauthor{\bsnm{Rosenman},~\bfnm{Evan}\binits{E.}},
  \bauthor{\bsnm{Basse},~\bfnm{Guillaume}\binits{G.}},
  \bauthor{\bsnm{Owen},~\bfnm{Art}\binits{A.}} \AND
  \bauthor{\bsnm{Baiocchi},~\bfnm{Michael}\binits{M.}}
(\byear{2020}).
\btitle{Combining Observational and Experimental Datasets Using Shrinkage
  Estimators}.
\end{barticle}
\endbibitem

\bibitem[\protect\citeauthoryear{Rothwell}{2005}]{rothwell2005}
\begin{barticle}[author]
\bauthor{\bsnm{Rothwell},~\bfnm{Peter~M}\binits{P.~M.}}
(\byear{2005}).
\btitle{External Validity of Randomised Controlled Trials: ``To Whom Do the
  Results of This Trial Apply?''}.
\bjournal{The Lancet}
\bvolume{365}
\bpages{82--93}.
\bdoi{10.1016/S0140-6736(04)17670-8}
\end{barticle}
\endbibitem

\bibitem[\protect\citeauthoryear{Rubin}{1974}]{rubin1974}
\begin{barticle}[author]
\bauthor{\bsnm{Rubin},~\bfnm{Donald~B.}\binits{D.~B.}}
(\byear{1974}).
\btitle{Estimating Causal Effects of Treatments in Randomized and Nonrandomized
  Studies.}
\bjournal{Journal of Educational Psychology}
\bvolume{66}
\bpages{688--701}.
\bdoi{10.1037/h0037350}
\end{barticle}
\endbibitem

\bibitem[\protect\citeauthoryear{Rudolph and {van Der
  Laan}}{2017}]{rudolph2017}
\begin{barticle}[author]
\bauthor{\bsnm{Rudolph},~\bfnm{Ke}\binits{K.}} \AND \bauthor{\bsnm{{van Der
  Laan}},~\bfnm{Mj}\binits{M.}}
(\byear{2017}).
\btitle{Robust Estimation of Encouragement Design Intervention Effects
  Transported across Sites}.
\bjournal{Journal Of The Royal Statistical Society Series B-Statistical
  Methodology}
\bvolume{79}
\bpages{1509--1525}.
\end{barticle}
\endbibitem

\bibitem[\protect\citeauthoryear{Schmid et~al.}{2020}]{schmid2020}
\begin{barticle}[author]
\bauthor{\bsnm{Schmid},~\bfnm{Ian}\binits{I.}},
  \bauthor{\bsnm{Rudolph},~\bfnm{Kara~E}\binits{K.~E.}},
  \bauthor{\bsnm{Nguyen},~\bfnm{Trang~Quynh}\binits{T.~Q.}},
  \bauthor{\bsnm{Hong},~\bfnm{Hwanhee}\binits{H.}},
  \bauthor{\bsnm{Seamans},~\bfnm{Marissa~J}\binits{M.~J.}},
  \bauthor{\bsnm{Ackerman},~\bfnm{Benjamin}\binits{B.}} \AND
  \bauthor{\bsnm{Stuart},~\bfnm{Elizabeth~A}\binits{E.~A.}}
(\byear{2020}).
\btitle{Comparing the Performance of Statistical Methods That Generalize Effect
  Estimates from Randomized Controlled Trials to Much Larger Target
  Populations}.
\bjournal{Communications in statistics. Simulation and computation}
\bvolume{ahead-of-print}
\bpages{1--23}.
\end{barticle}
\endbibitem

\bibitem[\protect\citeauthoryear{Schulz, Altman and Moher}{2010}]{schulz2010}
\begin{barticle}[author]
\bauthor{\bsnm{Schulz},~\bfnm{K.~F}\binits{K.~F.}},
  \bauthor{\bsnm{Altman},~\bfnm{D.~G}\binits{D.~G.}} \AND
  \bauthor{\bsnm{Moher},~\bfnm{D.}\binits{D.}}
(\byear{2010}).
\btitle{{{CONSORT}} 2010 Statement: Updated Guidelines for Reporting Parallel
  Group Randomised Trials}.
\bjournal{BMJ}
\bvolume{340}
\bpages{c332-c332}.
\bdoi{10.1136/bmj.c332}
\end{barticle}
\endbibitem

\bibitem[\protect\citeauthoryear{Schwartz and Lellouch}{1967}]{schwartz1967}
\begin{barticle}[author]
\bauthor{\bsnm{Schwartz},~\bfnm{Daniel}\binits{D.}} \AND
  \bauthor{\bsnm{Lellouch},~\bfnm{Joseph}\binits{J.}}
(\byear{1967}).
\btitle{Explanatory and Pragmatic Attitudes in Therapeutical Trials}.
\bjournal{Journal of Chronic Diseases}
\bvolume{20}
\bpages{637--648}.
\bdoi{10.1016/0021-9681(67)90041-0}
\end{barticle}
\endbibitem

\bibitem[\protect\citeauthoryear{Sen et~al.}{2016}]{sen2016}
\begin{barticle}[author]
\bauthor{\bsnm{Sen},~\bfnm{Anando}\binits{A.}},
  \bauthor{\bsnm{Chakrabarti},~\bfnm{Shreya}\binits{S.}},
  \bauthor{\bsnm{Goldstein},~\bfnm{Andrew}\binits{A.}},
  \bauthor{\bsnm{Wang},~\bfnm{Shuang}\binits{S.}},
  \bauthor{\bsnm{Ryan},~\bfnm{Patrick~B.}\binits{P.~B.}} \AND
  \bauthor{\bsnm{Weng},~\bfnm{Chunhua}\binits{C.}}
(\byear{2016}).
\btitle{{{GIST}} 2.0: A Scalable Multi-Trait Metric for Quantifying Population
  Representativeness of Individual Clinical Studies}.
\bjournal{Journal of Biomedical Informatics}
\bvolume{63}
\bpages{325--336}.
\bdoi{10.1016/j.jbi.2016.09.003}
\end{barticle}
\endbibitem

\bibitem[\protect\citeauthoryear{Shadish, Cook and
  Campbell}{2001}]{shadish2001}
\begin{bbook}[author]
\bauthor{\bsnm{Shadish},~\bfnm{William~R.}\binits{W.~R.}},
  \bauthor{\bsnm{Cook},~\bfnm{Thomas~D.}\binits{T.~D.}} \AND
  \bauthor{\bsnm{Campbell},~\bfnm{Donald~T.}\binits{D.~T.}}
(\byear{2001}).
\btitle{Experimental and Quasi-Experimental Designs for Generalized Causal
  Inference}.
\bpublisher{{Houghton Mifflin}}, \baddress{{Boston}}.
\end{bbook}
\endbibitem

\bibitem[\protect\citeauthoryear{Signorovitch et~al.}{2010}]{signorovitch2010}
\begin{barticle}[author]
\bauthor{\bsnm{Signorovitch},~\bfnm{James~E}\binits{J.~E.}},
  \bauthor{\bsnm{Wu},~\bfnm{Eric~Q}\binits{E.~Q.}},
  \bauthor{\bsnm{Yu},~\bfnm{Andrew~P}\binits{A.~P.}},
  \bauthor{\bsnm{Gerrits},~\bfnm{Charles~M}\binits{C.~M.}},
  \bauthor{\bsnm{Kantor},~\bfnm{Evan}\binits{E.}},
  \bauthor{\bsnm{Bao},~\bfnm{Yanjun}\binits{Y.}},
  \bauthor{\bsnm{Gupta},~\bfnm{Shiraz~R}\binits{S.~R.}} \AND
  \bauthor{\bsnm{Mulani},~\bfnm{Parvez~M}\binits{P.~M.}}
(\byear{2010}).
\btitle{Comparative Effectiveness without Head-to-Head Trials: A Method for
  Matching-Adjusted Indirect Comparisons Applied to Psoriasis Treatment with
  Adalimumab or Etanercept}.
\bjournal{PharmacoEconomics}
\bvolume{28}
\bpages{935--945}.
\end{barticle}
\endbibitem

\bibitem[\protect\citeauthoryear{Simon}{1982}]{simon1982}
\begin{barticle}[author]
\bauthor{\bsnm{Simon},~\bfnm{R.}\binits{R.}}
(\byear{1982}).
\btitle{Patient Subsets and Variation in Therapeutic Efficacy}.
\bjournal{British Journal of Clinical Pharmacology}
\bvolume{14}
\bpages{473--482}.
\bdoi{10.1111/j.1365-2125.1982.tb02015.x}
\end{barticle}
\endbibitem

\bibitem[\protect\citeauthoryear{Stuart}{2010}]{stuart2010}
\begin{barticle}[author]
\bauthor{\bsnm{Stuart},~\bfnm{Elizabeth~A.}\binits{E.~A.}}
(\byear{2010}).
\btitle{Matching Methods for Causal Inference: A Review and a Look Forward}.
\bjournal{Statistical Science}
\bvolume{25}
\bpages{1--21}.
\bdoi{10.1214/09-STS313}
\end{barticle}
\endbibitem

\bibitem[\protect\citeauthoryear{Stuart, Ackerman and
  Westreich}{2018}]{stuart2018}
\begin{barticle}[author]
\bauthor{\bsnm{Stuart},~\bfnm{Elizabeth~A.}\binits{E.~A.}},
  \bauthor{\bsnm{Ackerman},~\bfnm{Benjamin}\binits{B.}} \AND
  \bauthor{\bsnm{Westreich},~\bfnm{Daniel}\binits{D.}}
(\byear{2018}).
\btitle{Generalizability of Randomized Trial Results to Target Populations:
  Design and Analysis Possibilities}.
\bjournal{Research on Social Work Practice}
\bvolume{28}
\bpages{532--537}.
\bdoi{10.1177/1049731517720730}
\end{barticle}
\endbibitem

\bibitem[\protect\citeauthoryear{Stuart, Bradshaw and Leaf}{2015}]{stuart2015}
\begin{barticle}[author]
\bauthor{\bsnm{Stuart},~\bfnm{Elizabeth~A.}\binits{E.~A.}},
  \bauthor{\bsnm{Bradshaw},~\bfnm{Catherine~P.}\binits{C.~P.}} \AND
  \bauthor{\bsnm{Leaf},~\bfnm{Philip~J.}\binits{P.~J.}}
(\byear{2015}).
\btitle{Assessing the Generalizability of Randomized Trial Results to Target
  Populations}.
\bjournal{Prevention Science}
\bvolume{16}
\bpages{475--485}.
\bdoi{10.1007/s11121-014-0513-z}
\end{barticle}
\endbibitem

\bibitem[\protect\citeauthoryear{Stuart et~al.}{2011}]{stuart2011}
\begin{barticle}[author]
\bauthor{\bsnm{Stuart},~\bfnm{Elizabeth~A.}\binits{E.~A.}},
  \bauthor{\bsnm{Cole},~\bfnm{Stephen~R.}\binits{S.~R.}},
  \bauthor{\bsnm{Bradshaw},~\bfnm{Catherine~P.}\binits{C.~P.}} \AND
  \bauthor{\bsnm{Leaf},~\bfnm{Philip~J.}\binits{P.~J.}}
(\byear{2011}).
\btitle{The Use of Propensity Scores to Assess the Generalizability of Results
  from Randomized Trials: Use of Propensity Scores to Assess Generalizability}.
\bjournal{Journal of the Royal Statistical Society: Series A (Statistics in
  Society)}
\bvolume{174}
\bpages{369--386}.
\bdoi{10.1111/j.1467-985X.2010.00673.x}
\end{barticle}
\endbibitem

\bibitem[\protect\citeauthoryear{Su et~al.}{2008}]{su2008}
\begin{barticle}[author]
\bauthor{\bsnm{Su},~\bfnm{Xiaogang}\binits{X.}},
  \bauthor{\bsnm{Zhou},~\bfnm{Tianni}\binits{T.}},
  \bauthor{\bsnm{Yan},~\bfnm{Xin}\binits{X.}},
  \bauthor{\bsnm{Fan},~\bfnm{Juanjuan}\binits{J.}} \AND
  \bauthor{\bsnm{Yang},~\bfnm{Song}\binits{S.}}
(\byear{2008}).
\btitle{Interaction Trees with Censored Survival Data}.
\bjournal{The International Journal of Biostatistics}
\bvolume{4}.
\bdoi{10.2202/1557-4679.1071}
\end{barticle}
\endbibitem

\bibitem[\protect\citeauthoryear{Su et~al.}{2009}]{su2009}
\begin{barticle}[author]
\bauthor{\bsnm{Su},~\bfnm{Xg}\binits{X.}},
  \bauthor{\bsnm{Tsai},~\bfnm{Cl}\binits{C.}},
  \bauthor{\bsnm{Wang},~\bfnm{Hs}\binits{H.}},
  \bauthor{\bsnm{Nickerson},~\bfnm{DM}\binits{D.}} \AND
  \bauthor{\bsnm{Li},~\bfnm{Bg}\binits{B.}}
(\byear{2009}).
\btitle{Subgroup Analysis via Recursive Partitioning}.
\bjournal{Journal Of Machine Learning Research}
\bvolume{10}
\bpages{141--158}.
\end{barticle}
\endbibitem

\bibitem[\protect\citeauthoryear{Tian et~al.}{2014}]{tian2014}
\begin{barticle}[author]
\bauthor{\bsnm{Tian},~\bfnm{Lu}\binits{L.}},
  \bauthor{\bsnm{Alizadeh},~\bfnm{Ash~A.}\binits{A.~A.}},
  \bauthor{\bsnm{Gentles},~\bfnm{Andrew~J.}\binits{A.~J.}} \AND
  \bauthor{\bsnm{Tibshirani},~\bfnm{Robert}\binits{R.}}
(\byear{2014}).
\btitle{A Simple Method for Estimating Interactions between a Treatment and a
  Large Number of Covariates}.
\bjournal{Journal of the American Statistical Association}
\bvolume{109}
\bpages{1517--1532}.
\bdoi{10.1080/01621459.2014.951443}
\end{barticle}
\endbibitem

\bibitem[\protect\citeauthoryear{Tipton}{2013a}]{tipton2013a}
\begin{barticle}[author]
\bauthor{\bsnm{Tipton},~\bfnm{Elizabeth}\binits{E.}}
(\byear{2013}a).
\btitle{Stratified Sampling Using Cluster Analysis: A Sample Selection Strategy
  for Improved Generalizations from Experiments}.
\bjournal{Evaluation Review}
\bvolume{37}
\bpages{109--139}.
\bdoi{10.1177/0193841X13516324}
\end{barticle}
\endbibitem

\bibitem[\protect\citeauthoryear{Tipton}{2013b}]{tipton2013}
\begin{barticle}[author]
\bauthor{\bsnm{Tipton},~\bfnm{Elizabeth}\binits{E.}}
(\byear{2013}b).
\btitle{Improving Generalizations from Experiments Using Propensity Score
  Subclassification: Assumptions, Properties, and Contexts}.
\bjournal{Journal of Educational and Behavioral Statistics}
\bvolume{38}
\bpages{239--266}.
\bdoi{10.3102/1076998612441947}
\end{barticle}
\endbibitem

\bibitem[\protect\citeauthoryear{Tipton}{2014}]{tipton2014}
\begin{barticle}[author]
\bauthor{\bsnm{Tipton},~\bfnm{Elizabeth}\binits{E.}}
(\byear{2014}).
\btitle{How Generalizable Is Your Experiment? {{An}} Index for Comparing
  Experimental Samples and Populations}.
\bjournal{Journal of Educational and Behavioral Statistics}
\bvolume{39}
\bpages{478--501}.
\bdoi{10.3102/1076998614558486}
\end{barticle}
\endbibitem

\bibitem[\protect\citeauthoryear{Tipton and Olsen}{2018}]{tipton2018}
\begin{barticle}[author]
\bauthor{\bsnm{Tipton},~\bfnm{Elizabeth}\binits{E.}} \AND
  \bauthor{\bsnm{Olsen},~\bfnm{Robert~B.}\binits{R.~B.}}
(\byear{2018}).
\btitle{A Review of Statistical Methods for Generalizing from Evaluations of
  Educational Interventions}.
\bjournal{Educational Researcher}
\bvolume{47}
\bpages{516--524}.
\bdoi{10.3102/0013189X18781522}
\end{barticle}
\endbibitem

\bibitem[\protect\citeauthoryear{Tipton and Peck}{2017}]{tipton2017}
\begin{barticle}[author]
\bauthor{\bsnm{Tipton},~\bfnm{Elizabeth}\binits{E.}} \AND
  \bauthor{\bsnm{Peck},~\bfnm{Laura~R.}\binits{L.~R.}}
(\byear{2017}).
\btitle{A Design-Based Approach to Improve External Validity in Welfare Policy
  Evaluations}.
\bjournal{Evaluation Review}
\bvolume{41}
\bpages{326--356}.
\bdoi{10.1177/0193841X16655656}
\end{barticle}
\endbibitem

\bibitem[\protect\citeauthoryear{Tipton et~al.}{2014}]{tipton2014a}
\begin{barticle}[author]
\bauthor{\bsnm{Tipton},~\bfnm{Elizabeth}\binits{E.}},
  \bauthor{\bsnm{Hedges},~\bfnm{Larry}\binits{L.}},
  \bauthor{\bsnm{{Vaden-Kiernan}},~\bfnm{Michael}\binits{M.}},
  \bauthor{\bsnm{Borman},~\bfnm{Geoffrey}\binits{G.}},
  \bauthor{\bsnm{Sullivan},~\bfnm{Kate}\binits{K.}} \AND
  \bauthor{\bsnm{Caverly},~\bfnm{Sarah}\binits{S.}}
(\byear{2014}).
\btitle{Sample Selection in Randomized Experiments: A New Method Using
  Propensity Score Stratified Sampling}.
\bjournal{Journal of Research on Educational Effectiveness}
\bvolume{7}
\bpages{114--135}.
\bdoi{10.1080/19345747.2013.831154}
\end{barticle}
\endbibitem

\bibitem[\protect\citeauthoryear{Tipton et~al.}{2017}]{tipton2017a}
\begin{barticle}[author]
\bauthor{\bsnm{Tipton},~\bfnm{Elizabeth}\binits{E.}},
  \bauthor{\bsnm{Hallberg},~\bfnm{Kelly}\binits{K.}},
  \bauthor{\bsnm{Hedges},~\bfnm{Larry~V.}\binits{L.~V.}} \AND
  \bauthor{\bsnm{Chan},~\bfnm{Wendy}\binits{W.}}
(\byear{2017}).
\btitle{Implications of Small Samples for Generalization: Adjustments and Rules
  of Thumb}.
\bjournal{Evaluation Review}
\bvolume{41}
\bpages{472--505}.
\bdoi{10.1177/0193841X16655665}
\end{barticle}
\endbibitem

\bibitem[\protect\citeauthoryear{Turner et~al.}{2009}]{turner2009}
\begin{barticle}[author]
\bauthor{\bsnm{Turner},~\bfnm{Rebecca~M.}\binits{R.~M.}},
  \bauthor{\bsnm{Spiegelhalter},~\bfnm{David~J.}\binits{D.~J.}},
  \bauthor{\bsnm{Smith},~\bfnm{Gordon C.~S.}\binits{G.~C.~S.}} \AND
  \bauthor{\bsnm{Thompson},~\bfnm{Simon~G.}\binits{S.~G.}}
(\byear{2009}).
\btitle{Bias Modelling in Evidence Synthesis}.
\bjournal{Journal of the Royal Statistical Society: Series A (Statistics in
  Society)}
\bvolume{172}
\bpages{21--47}.
\end{barticle}
\endbibitem

\bibitem[\protect\citeauthoryear{{Van der Laan}, Laan and
  Robins}{2003}]{vanderlaan2003}
\begin{bbook}[author]
\bauthor{\bsnm{{Van der Laan}},~\bfnm{Mark~J}\binits{M.~J.}},
  \bauthor{\bsnm{Laan},~\bfnm{MJ}\binits{M.}} \AND
  \bauthor{\bsnm{Robins},~\bfnm{James~M}\binits{J.~M.}}
(\byear{2003}).
\btitle{Unified Methods for Censored Longitudinal Data and Causality}.
\bpublisher{{Springer Science \& Business Media}}.
\end{bbook}
\endbibitem

\bibitem[\protect\citeauthoryear{{van der Laan} and
  Rose}{2011}]{vanderlaan2011}
\begin{bbook}[author]
\bauthor{\bsnm{{van der Laan}},~\bfnm{Mark~J.}\binits{M.~J.}} \AND
  \bauthor{\bsnm{Rose},~\bfnm{Sherri}\binits{S.}}
(\byear{2011}).
\btitle{Targeted Learning}.
\bseries{Springer {{Series}} in {{Statistics}}}.
\bpublisher{{Springer New York}}, \baddress{{New York, NY}}.
\bdoi{10.1007/978-1-4419-9782-1}
\end{bbook}
\endbibitem

\bibitem[\protect\citeauthoryear{Varadhan, Henderson and
  Weiss}{2016}]{varadhan2016}
\begin{barticle}[author]
\bauthor{\bsnm{Varadhan},~\bfnm{Ravi}\binits{R.}},
  \bauthor{\bsnm{Henderson},~\bfnm{Nicholas~C.}\binits{N.~C.}} \AND
  \bauthor{\bsnm{Weiss},~\bfnm{Carlos~O.}\binits{C.~O.}}
(\byear{2016}).
\btitle{Cross-Design Synthesis for Extending the Applicability of Trial
  Evidence When Treatment Effect Is Heterogeneous: Part i. Methodology}.
\bjournal{Communications in Statistics: Case Studies, Data Analysis and
  Applications}
\bvolume{2}
\bpages{112--126}.
\bdoi{10.1080/23737484.2017.1392265}
\end{barticle}
\endbibitem

\bibitem[\protect\citeauthoryear{Verde}{2019}]{verde2019}
\begin{barticle}[author]
\bauthor{\bsnm{Verde},~\bfnm{Pablo~Emilio}\binits{P.~E.}}
(\byear{2019}).
\btitle{The Hierarchical Metaregression Approach and Learning from Clinical
  Evidence}.
\bjournal{Biometrical Journal}
\bvolume{61}
\bpages{535--557}.
\bdoi{10.1002/bimj.201700266}
\end{barticle}
\endbibitem

\bibitem[\protect\citeauthoryear{Verde and Ohmann}{2015}]{verde2015}
\begin{barticle}[author]
\bauthor{\bsnm{Verde},~\bfnm{Pablo~E.}\binits{P.~E.}} \AND
  \bauthor{\bsnm{Ohmann},~\bfnm{Christian}\binits{C.}}
(\byear{2015}).
\btitle{Combining Randomized and Non-Randomized Evidence in Clinical Research:
  A Review of Methods and Applications: Combining Randomized and Non-Randomized
  Evidence}.
\bjournal{Research Synthesis Methods}
\bvolume{6}
\bpages{45--62}.
\bdoi{10.1002/jrsm.1122}
\end{barticle}
\endbibitem

\bibitem[\protect\citeauthoryear{Verde et~al.}{2016}]{verde2016}
\begin{barticle}[author]
\bauthor{\bsnm{Verde},~\bfnm{Pablo~E.}\binits{P.~E.}},
  \bauthor{\bsnm{Ohmann},~\bfnm{Christian}\binits{C.}},
  \bauthor{\bsnm{Morbach},~\bfnm{Stephan}\binits{S.}} \AND
  \bauthor{\bsnm{Icks},~\bfnm{Andrea}\binits{A.}}
(\byear{2016}).
\btitle{Bayesian Evidence Synthesis for Exploring Generalizability of Treatment
  Effects: A Case Study of Combining Randomized and Non-Randomized Results in
  Diabetes: Bayesian Evidence Synthesis for Exploring Generalizability of
  Treatment Effects: A Case Study of Combining Randomized and Non-Randomized
  Results in Di}.
\bjournal{Statistics in Medicine}
\bvolume{35}
\bpages{1654--1675}.
\bdoi{10.1002/sim.6809}
\end{barticle}
\endbibitem

\bibitem[\protect\citeauthoryear{{von Elm} et~al.}{2008}]{vonelm2008}
\begin{barticle}[author]
\bauthor{\bsnm{{von Elm}},~\bfnm{Erik}\binits{E.}},
  \bauthor{\bsnm{Altman},~\bfnm{Douglas~G.}\binits{D.~G.}},
  \bauthor{\bsnm{Egger},~\bfnm{Matthias}\binits{M.}},
  \bauthor{\bsnm{Pocock},~\bfnm{Stuart~J.}\binits{S.~J.}},
  \bauthor{\bsnm{G{\o}tzsche},~\bfnm{Peter~C.}\binits{P.~C.}} \AND
  \bauthor{\bsnm{Vandenbroucke},~\bfnm{Jan~P.}\binits{J.~P.}}
(\byear{2008}).
\btitle{The Strengthening the Reporting of Observational Studies in
  Epidemiology ({{STROBE}}) Statement: Guidelines for Reporting Observational
  Studies}.
\bjournal{Journal of Clinical Epidemiology}
\bvolume{61}
\bpages{344--349}.
\bdoi{10.1016/j.jclinepi.2007.11.008}
\end{barticle}
\endbibitem

\bibitem[\protect\citeauthoryear{Weisberg, Hayden and
  Pontes}{2009}]{weisberg2009}
\begin{barticle}[author]
\bauthor{\bsnm{Weisberg},~\bfnm{Herbert~I}\binits{H.~I.}},
  \bauthor{\bsnm{Hayden},~\bfnm{Vanessa~C}\binits{V.~C.}} \AND
  \bauthor{\bsnm{Pontes},~\bfnm{Victor~P}\binits{V.~P.}}
(\byear{2009}).
\btitle{Selection Criteria and Generalizability within the Counterfactual
  Framework: Explaining the Paradox of Antidepressant-Induced Suicidality?}
\bjournal{Clinical Trials}
\bvolume{6}
\bpages{109--18}.
\bdoi{10.1177/1740774509102563}
\end{barticle}
\endbibitem

\bibitem[\protect\citeauthoryear{Weiss, Segal and Varadhan}{2012}]{weiss2012}
\begin{barticle}[author]
\bauthor{\bsnm{Weiss},~\bfnm{Carlos~O.}\binits{C.~O.}},
  \bauthor{\bsnm{Segal},~\bfnm{Jodi~B.}\binits{J.~B.}} \AND
  \bauthor{\bsnm{Varadhan},~\bfnm{Ravi}\binits{R.}}
(\byear{2012}).
\btitle{Assessing the Applicability of Trial Evidence to a Target Sample in the
  Presence of Heterogeneity of Treatment Effect: {{APPLICABILITY OF TREATMENT
  EFFECTS}}}.
\bjournal{Pharmacoepidemiology and Drug Safety}
\bvolume{21}
\bpages{121--129}.
\bdoi{10.1002/pds.3242}
\end{barticle}
\endbibitem

\bibitem[\protect\citeauthoryear{Weng et~al.}{2014}]{weng2014}
\begin{barticle}[author]
\bauthor{\bsnm{Weng},~\bfnm{C}\binits{C.}},
  \bauthor{\bsnm{Li},~\bfnm{Y}\binits{Y.}},
  \bauthor{\bsnm{Ryan},~\bfnm{P}\binits{P.}},
  \bauthor{\bsnm{Zhang},~\bfnm{Y}\binits{Y.}},
  \bauthor{\bsnm{Liu},~\bfnm{F}\binits{F.}},
  \bauthor{\bsnm{Gao},~\bfnm{J}\binits{J.}},
  \bauthor{\bsnm{Bigger},~\bfnm{J.~T}\binits{J.~T.}} \AND
  \bauthor{\bsnm{Hripcsak},~\bfnm{G}\binits{G.}}
(\byear{2014}).
\btitle{A Distribution-Based Method for Assessing the Differences between
  Clinical Trial Target Populations and Patient Populations in Electronic
  Health Records}.
\bjournal{Applied clinical informatics}
\bvolume{5}
\bpages{463--479}.
\end{barticle}
\endbibitem

\bibitem[\protect\citeauthoryear{Westreich et~al.}{2017}]{westreich2017}
\begin{barticle}[author]
\bauthor{\bsnm{Westreich},~\bfnm{Daniel}\binits{D.}},
  \bauthor{\bsnm{Edwards},~\bfnm{Jessie~K}\binits{J.~K.}},
  \bauthor{\bsnm{Lesko},~\bfnm{Catherine~R}\binits{C.~R.}},
  \bauthor{\bsnm{Stuart},~\bfnm{Elizabeth}\binits{E.}} \AND
  \bauthor{\bsnm{Cole},~\bfnm{Stephen~R}\binits{S.~R.}}
(\byear{2017}).
\btitle{Transportability of Trial Results Using Inverse Odds of Sampling
  Weights}.
\bjournal{American Journal of Epidemiology}
\bvolume{186}
\bpages{1010--1014}.
\bdoi{10.1093/aje/kwx164}
\end{barticle}
\endbibitem

\end{thebibliography}

\pagebreak
\appendix

\section*{Appendix: Summary of methods that only require summary-level data}

Without access to individual patient data in the study and/or target samples, investigators will be constrained as to the estimators available to them. The following estimators can be applied in this setting. Investigators should strive to maximally use the available data and hence use methods that incorporate individual-level data where they are available. 

\paragraph{Summary-level data for both study (covariate and outcome) and target samples (covariate).} Post-stratification \citep{miettinen1972,prentice2005} only requires joint distributions or cell counts for each stratum. Using only study and target sample means, one could also apply outcome regressions that are linear in their predictors.

\paragraph{Summary-level outcome data for both study and target samples.} Bias-adjusted meta-analysis approaches by \cite{turner2009} and \cite{greenland2005} require summary-level study outcome data with estimates of bias for each study. When that summary-level data are stratified by effect modifiers, one can use approaches by \cite{eddy1989} and \cite{prevost2000}.  If summary-level study data are stratified by participants included vs. excluded from the study, cross-design synthesis can be used \citep{begg1992,kaizar2011}.

\paragraph{Summary-level covariate and outcome data in the study, individual-level covariate and outcome data in the target sample.} With summary-level study and individual-level target sample data, one can use hierarchical Bayesian evidence synthesis \citep{verde2016,verde2019}.

\paragraph{Individual-level covariate and outcome data in the study, summary-level covariate data in the target sample.} With individual-level study and summary-level target data, one can use matching with reweighting (e.g., \cite{hartman2015}), or \cite{signorovitch2010} or \cite{phillippo2018}'s propensity and outcome regression approaches. When joint distributions of summary-level target sample data are available, one can use IPPW \citep{cole2010,westreich2017}.

\end{document}